\documentclass[11pt]{article}
\usepackage{jmlr2e}
\usepackage{abstract}
\usepackage{array}
\usepackage{bbding}
\usepackage{multirow}
\usepackage{adjustbox}
\usepackage{bbm}
\usepackage{float}
\usepackage{amsfonts}
\usepackage{graphicx}
\graphicspath{{graphics/}}
\usepackage{subfigure}
\usepackage{float}

\usepackage{tabularx,booktabs}
\usepackage{multirow}
\newcolumntype{Y}{>{\centering\arraybackslash}X}

\usepackage[ruled,vlined,algo2e]{algorithm2e}
\usepackage{algorithm}
\usepackage{algorithmic}
\usepackage{natbib}

\usepackage{lscape}
\usepackage{comment}
\usepackage{bm}
\usepackage[title]{appendix}
\usepackage{longtable}
\usepackage{mathtools}
\usepackage{dsfont}
\usepackage{fancyhdr}

\usepackage{stackengine}
\usepackage{textcomp}
\usepackage{stfloats}
\usepackage{verbatim}
\usepackage{xcolor}
\definecolor{darkblue}{rgb}{0.0,0.5,0.5}
\definecolor{blue}{rgb}{0.0,0.0,1}

\usepackage[colorlinks]{hyperref}
\hypersetup{colorlinks,breaklinks,linkcolor=blue,urlcolor=blue,anchorcolor=blue,citecolor=blue}
\usepackage{cleveref}
\usepackage{booktabs,caption}
\usepackage{multirow}
\usepackage{lscape}
\usepackage{epstopdf}
\usepackage{lineno}
\usepackage{microtype}

\usepackage{graphicx}
\usepackage{subfigure}
\usepackage{booktabs} 
\usepackage{bbm}

\usepackage{amsfonts}
\usepackage{xurl}
\usepackage{stackengine}
\usepackage{tikz}
\usetikzlibrary{decorations.pathreplacing}
\usetikzlibrary{positioning,arrows.meta,quotes}
\usetikzlibrary{shapes,snakes}
\usetikzlibrary{bayesnet}
\tikzset{>=latex}
\tikzstyle{plate caption} = [caption, node distance=0, inner sep=0pt, below left=5pt and 0pt of #1.south]
\usepackage[normalem]{ulem}
\usepackage{multirow}

\firstpageno{1}

\begin{document}

\title{Conditional forecasting of bus travel time and passenger occupancy with Bayesian Markov regime-switching vector autoregression}

\author{\name Xiaoxu Chen \email xiaoxu.chen@mail.mcgill.ca\\
       \addr Department of Civil Engineering\\
       McGill University\\
       Montreal, QC  H3A 0C3, Canada
       \AND
       \name Zhanhong Cheng \email zhanhong.cheng@mcgill.ca \\
       \addr Department of Civil Engineering\\
       McGill University\\
       Montreal, QC  H3A 0C3, Canada
       \AND
       \name Alexandra M. Schmidt \email alexandra.schmidt@mcgill.ca \\
       \addr Department of Epidemiology, Biostatistics and Occupational Health\\
       McGill University\\
       Montreal, QC  H3A 0C3, Canada
       \AND
       \name Lijun Sun\thanks{Corresponding author.} \email lijun.sun@mcgill.ca \\
       \addr Department of Civil Engineering\\
       McGill University\\
       Montreal, QC  H3A 0C3, Canada}

\editor{}

\maketitle

\begin{abstract}
Accurately forecasting bus travel time and passenger occupancy with uncertainty is essential for both travelers and transit agencies/operators. However, existing approaches to forecasting bus travel time and passenger occupancy mainly rely on deterministic models, providing only point estimates. In this paper, we develop a Bayesian Markov regime-switching vector autoregressive model to jointly forecast both bus travel time and passenger occupancy with uncertainty. The proposed approach naturally captures the intricate interactions among adjacent buses and adapts to the multimodality and skewness of real-world bus travel time and passenger occupancy observations. We develop an efficient Markov chain Monte Carlo (MCMC) sampling algorithm to approximate the resultant joint posterior distribution of the parameter vector. With this framework, the estimation of downstream bus travel time and passenger occupancy is transformed into a multivariate time series forecasting problem conditional on partially observed outcomes. Experimental validation using real-world data demonstrates the superiority of our proposed model in terms of both predictive means and uncertainty quantification compared to the Bayesian Gaussian mixture model.

\end{abstract}


\section{Introduction}
\label{sec:intro}
The rapid progress of urbanization has brought increasing population and economic agglomeration in large cities. Urban transportation problems such as increased traffic congestion and pollution, high energy consumption and greenhouse gas emissions, and growing safety and accessibility concerns, have been persistently challenging the development of sustainable cities and communities. In the \textit{2030 Agenda for Sustainable Development}, the United Nations has emphasized the critical role of public transportation in shaping a sustainable society \citep{resolution2015transforming}. However, despite the growing investment in infrastructure, North American cities have not seen rapid growth and even observed a decline in ridership in recent years, even before the COVID-19 pandemic \citep{erhardt2022has}. One of the key reasons is that the operation of transit services suffers from reliability issues. Bus operation is a highly challenging problem due to the inherent instability of the system---a slightly delayed bus will be further delayed as it will encounter more waiting passengers and experience longer dwell time. Unreliability in travel time and overcrowdedness resulting from unstable operations \citep[e.g., ``bus bunching'', see ][]{daganzo2009headway,bartholdi2012self} have been the main factors preventing travelers from using public transportation \citep{carrel2013passengers}. Having access to accurate travel time and occupancy forecasting along with uncertainty becomes important to travelers to make informed travel planning in terms of mode choice, route choice, and even vehicle choice (e.g., waiting for a less crowded bus or boarding a full vehicle) \citep{yu2017using}. For transit agencies/operators, probabilistic forecasting could benefit the design of robust bus management strategies, such as bus route design \citep[e.g.,][]{zheng2016reliable}, bus crowding control \citep[e.g.,][]{wang2021providing}, timetable design \citep[e.g.,][]{jiang2021optimal}, and bus bunching control \citep[e.g.,][]{xuan2011dynamic,bartholdi2012self}.

In general, a bus link refers to the one-way segment that connects two adjacent bus stops along a bus route. Link travel time of a bus is defined as the time difference between the arrivals at two adjacent bus stops associated with the bus link. Therefore, the trip travel time of the bus from one stop to another can be calculated by summing up all link travel times between these two stops. The passenger occupancy of a bus on the link is defined as the total number of passengers onboard while the bus is traversing that particular link. The main goal of this paper is to provide real-time predictions of downstream travel time and passenger occupancy of a bus along a given route.

Previous studies on forecasting bus travel time have predominantly employed deterministic approaches, based on techniques such as Artificial Neural Network \citep{gurmu2014artificial}, Support Vector Machine \citep{yu2011bus,kumar2013pattern}, K-nearest
neighbors model \citep{kumar2019real}, Long-Short-Term Memory neural network \citep{he2018travel}, and various hybrid models \citep{yu2018prediction}. In addition,  deterministic passenger occupancy forecasting models have also been developed, including Lasso regularized linear regression model \citep{jenelius2019data}, partial least squares regression \citep{jenelius2020personalized}, Random Forest \citep{wood2022development}, and deep learning model \citep{bapaume2023forecasting}. Despite the widespread use and simplicity of these deterministic models, a significant drawback is that they only provide point estimates, overlooking the randomness and uncertainty associated with the prediction.

There are only a few studies on probabilistic forecasting for bus travel time and passenger occupancy \citep[e.g.,][]{ma2017estimation,dai2019bus,buchel2022b,buchel2022a,chen2022probabilistic}. However, these studies often adopt oversimplified assumptions and ignore important operational characteristics of bus systems, including:
\begin{itemize}
    \item \textit{Strong interactions/correlations between travel time and passenger occupancy}. Previous studies on probabilistic prediction for bus travel time \citep{ma2017estimation,dai2019bus,buchel2022b,buchel2022a} and passenger occupancy \citep{wang2021two} have typically treated these variables independently. However, the intricate interactions and correlations between bus travel time and passenger occupancy are not adequately considered. For example, the boarding and alighting processes take longer for a crowded bus compared to an empty one \citep{sun2014models}. Incorporating these interactions through a joint modeling approach could significantly enhance the accuracy of probabilistic forecasting.

    \item \textit{Complex link correlations of both travel time and passenger occupancy}. \cite{chen2022bayesian} demonstrated that link travel times on a bus route exhibit complex local and long-range correlations. In Section~\ref{sec:data}, we also find that link passenger occupancy on a bus route exhibits local and long-range correlations. However, existing studies often focus only on local correlations while neglecting long-range correlations in probabilistic forecasting for both bus travel time \citep{ma2017estimation,buchel2022b,buchel2022a} and passenger occupancy \citep{wang2021two}.

    \item \textit{Interactions/correlations between adjacent buses along a bus route}. Adjacent buses often have strong interactions that lead to system instability and bus bunching  \citep{daganzo2009headway,bartholdi2012self}; for instance, due to the increase in headway (i.e., duration between two arrivals), a delayed bus could see more passengers waiting at the bus stop and get further delayed. To our knowledge, only \citet{buchel2022a} and \citet{chen2022probabilistic} considered the interactions/correlations between adjacent buses along a bus route in their proposed models for bus travel time forecasting, while most studies \citep{ma2017estimation,buchel2022b} focused on modeling the correlations between adjacent links along a bus route.
\end{itemize}

The recent contribution by \citet{chen2022probabilistic} develops a time-dependent Gaussian mixture model, which treats the concatenation of link travel time data from two consecutive buses as a random variable. The estimated model can then be used to perform conditional forecasting for the travel times of all downstream links. The study demonstrates that incorporating information from neighboring buses significantly improves forecasting accuracy. However, a notable limitation of this model is that two consecutive observations are assumed to be conditionally independent given their latent states, thus neglecting the temporal/dynamic relationships among multiple buses. This paper addresses the aforementioned challenges by developing a \textit{joint} Bayesian model for bus travel time and passenger occupancy, building upon the foundation laid by \citet{chen2022probabilistic}. To model the correlations between travel time and passenger occupancy, we construct a variable that combines the link travel time vector, the passenger occupancy vector, and the departure headway. More importantly, we employ a Bayesian Markov regime-switching vector autoregressive model to characterize the dynamic relationship among multiple buses. This new approach effectively captures essential interactions between adjacent buses, along with the multimodality and skewness of bus travel time and passenger occupancy distributions. Furthermore, it adeptly models intricate state transitions, particularly crucial when forecasting bus travel time and passenger occupancy with limited observations for the following bus. For model estimation, we develop an efficient Markov chain Monte Carlo (MCMC) algorithm to draw samples from the resulting posterior distribution of the model parameters. As we follow the Bayesian paradigm to estimate the parameters of the model, predictions are obtained by approximating the posterior predictive distribution. We fit the proposed model to the smart card data of one bus route in an anonymous city. The experimental results confirm that the proposed Markov regime-switching vector autoregressive model outperforms existing methods in terms of both point estimates and uncertainty quantification. This holistic approach contributes to a more robust and nuanced understanding of bus travel time and passenger occupancy dynamics, offering improved forecasting capabilities in real-world scenarios.

The remainder of this paper is organized as follows. In Section~\ref{sec:data}, we perform an empirical analysis of bus travel time and passenger occupancy using real-world data. In Section~\ref{sec:model}, we present the proposed Bayesian Markov regime-switching vector autoregressive model. The forecasting process and the probabilistic forecasting model are described in Section~\ref{sec:forecast}. We then showcase the capabilities of our proposed model by analyzing real-world data in Section~\ref{sec:exp}. Finally, we conclude the study and summarize our key findings in Section~\ref{sec:con}.

\section{Data description and analysis}
\label{sec:data}

Smart card data, a prevalent data source in public transit studies, is collected from electronic fare payment systems---transactions are created when passengers use smart cards (e.g., contactless cards or mobile payment apps) to pay for their bus trips \citep{pelletier2011smart}. These data sets capture information about boarding and alighting, including time, location, fare paid, and card ID. In this paper, we use the smart card data for one bus route (32 stops and 31 links) in an anonymous city to prepare the bus travel time and passenger occupancy data.

\begin{figure}[!t]
\centering
\includegraphics[width = 0.85\textwidth]{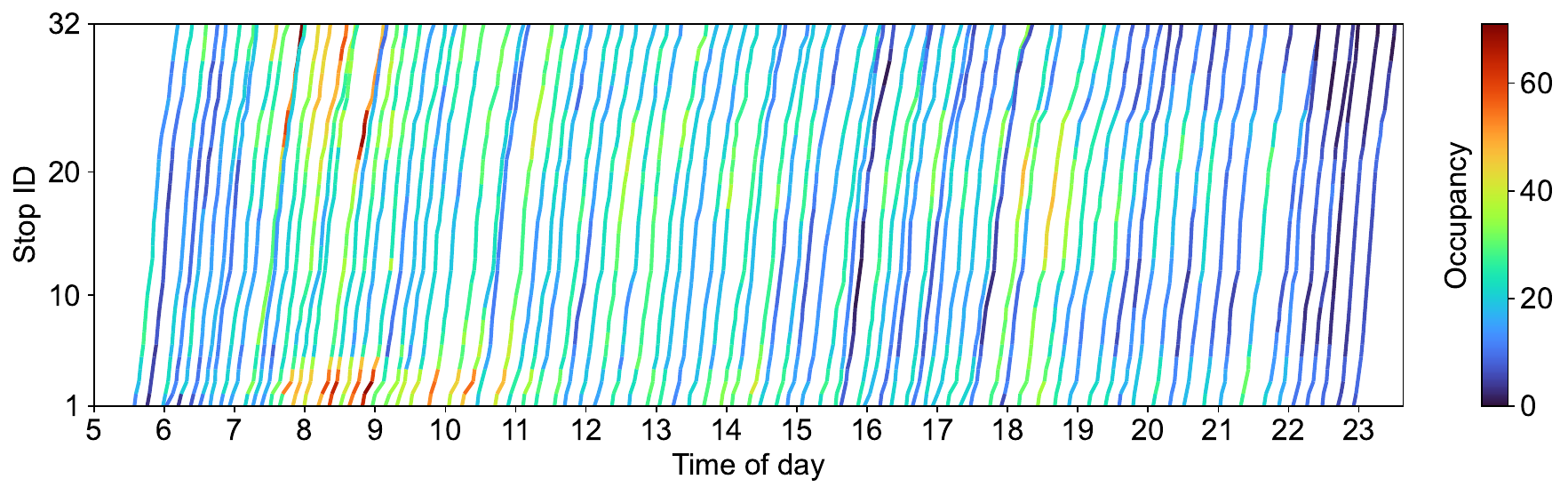}
\caption{Bus time-space trajectories with passenger occupancy. The horizontal axis represents the ``Time of day'' (from 5:00 to 24:00) and the vertical axis represents ``Stop ID'' (from the starting stop to the final stop, 32 stops in total). Each curve contains 31 segments, and it depicts the trajectory of a bus traveling from the departure stop to the final stop. The color of a segment shows the passenger occupancy of the bus.}
\label{fig:trajectory}
\end{figure}

Fig.~\ref{fig:trajectory} shows bus trajectories with passenger occupancy in one day. We can see that some adjacent buses have strong interactions like bus bunching, especially during morning peak hours. Bus bunching typically arises due to various factors, including traffic congestion, unpredictable passenger boarding and alighting times, variations in travel speeds, and delays caused by external factors such as road conditions, traffic signals, and accidents. When a bus falls behind schedule, it tends to experience increased passenger occupancy at subsequent stops and becomes further delayed. As a result, bus travel time and passenger occupancy have strong correlations, and adjacent buses on the same route often demonstrate complex interactions.

\begin{figure*}[!t]
\centering
\includegraphics[width = 0.9\textwidth]{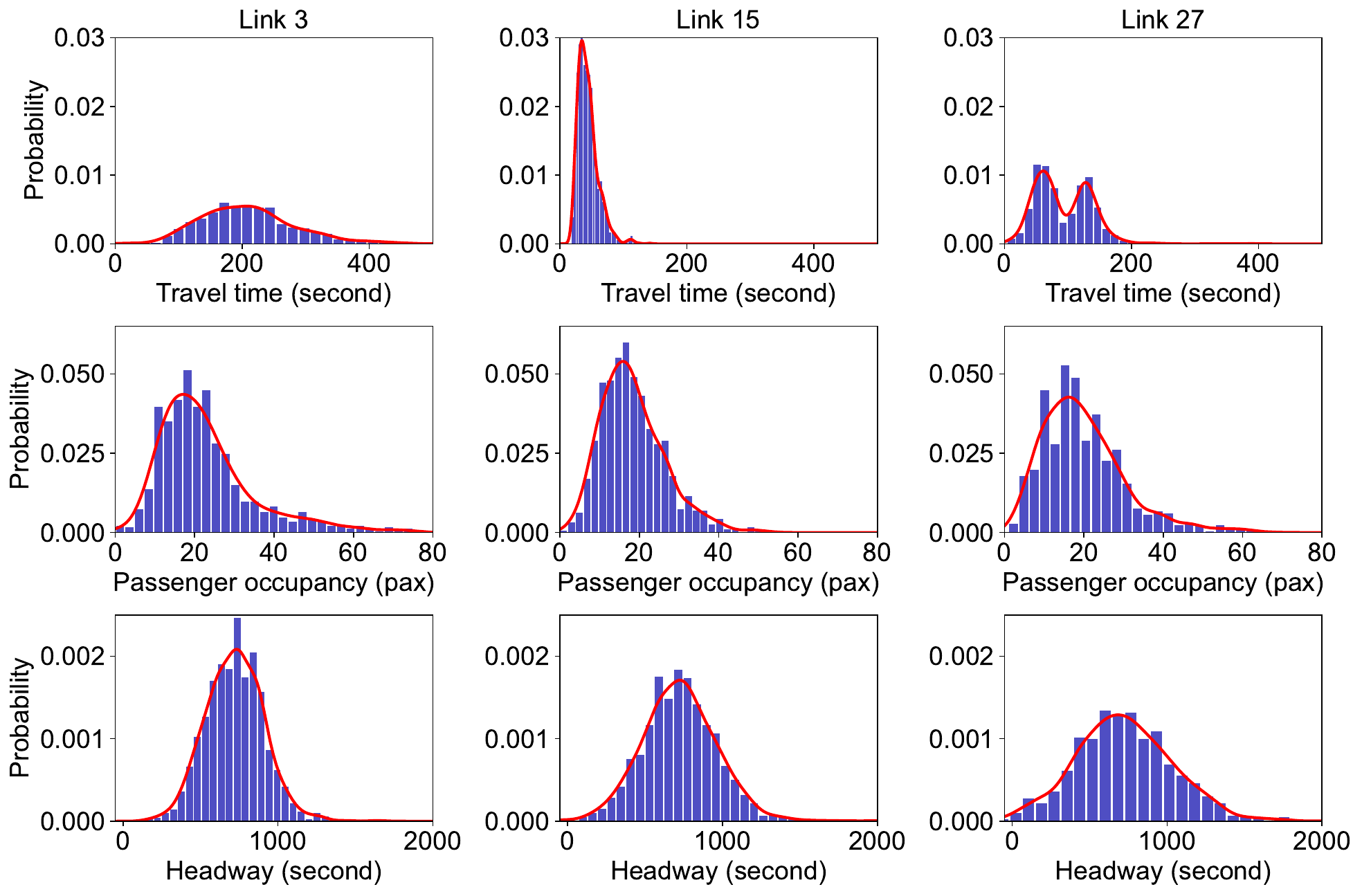}
\caption{Examples of empirical distributions of link travel time (first row), passenger occupancy (second row), and headway (third row). The red lines represent kernel density estimates of the empirical distributions across the links.}
\label{fig:distribution}
\end{figure*}

Fig.~\ref{fig:distribution} shows some example empirical distributions of travel time, passenger occupancy, and headway on three links. We observe clear characteristics such as positive skewness, heavy tails, and multimodality. Therefore, it is essential to develop a model that can effectively characterize such complex distributions to ensure the quality of probabilistic forecasting.

\begin{figure}[!t]
\centering
\includegraphics[width = 0.9\textwidth]{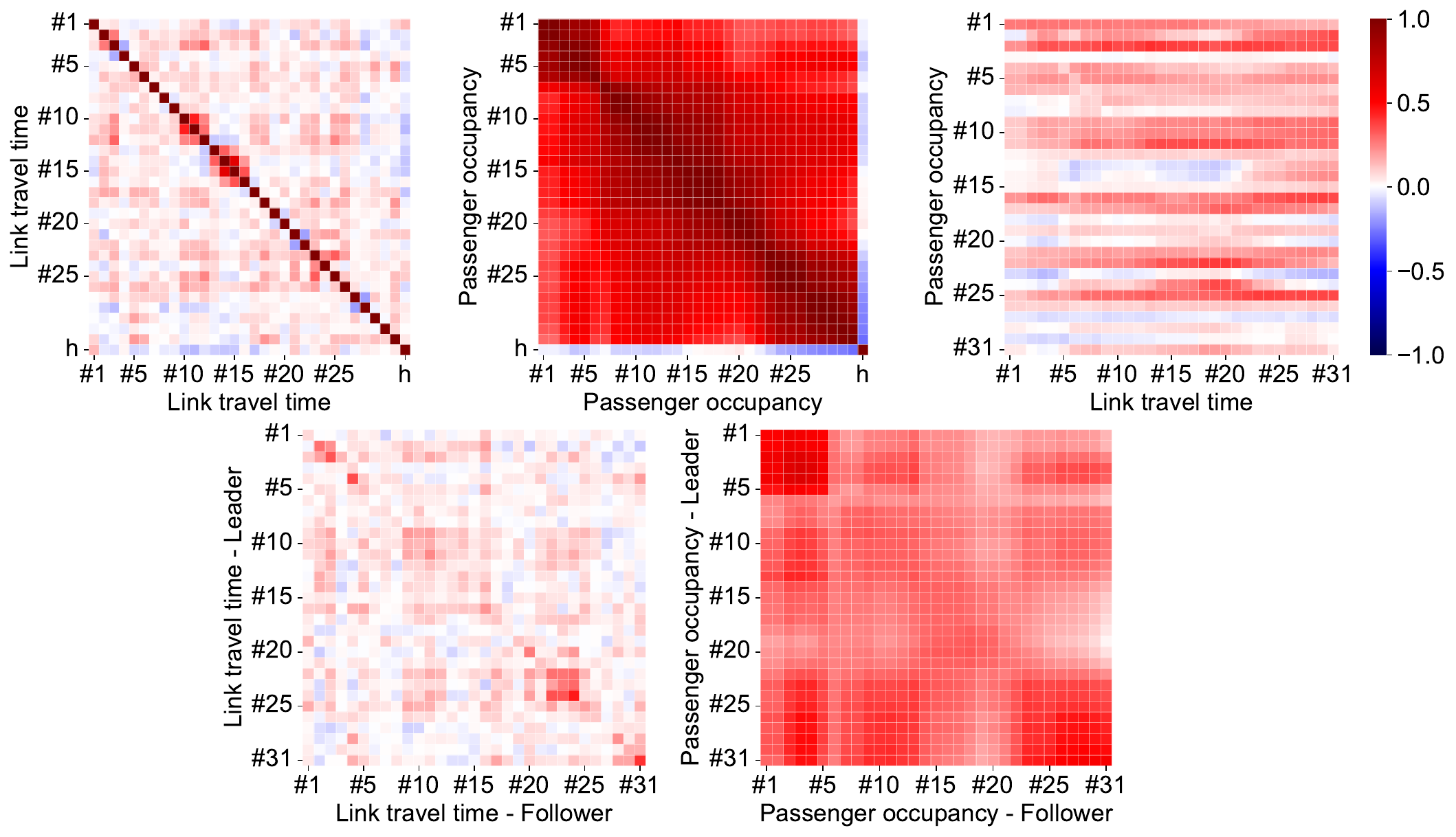}
\caption{Correlation and cross-correlation matrices of associated variables.}
\label{fig:correlation}
\end{figure}

We empirically compute the correlation and cross-correlation matrices among the associate variables, which are shown in Fig.~\ref{fig:correlation}. The first panel presents the link travel time correlation matrix, which reveals that link travel times have both local and long-range correlations. It is worth noting that the label $h$ in this matrix refers to the headway at the departure/originating stop, which enables us to capture the relationship between the link travel times and departure headway. Therefore, we could use observed link travel times and the headway to forecast downstream link travel times. The second panel presents the passenger occupancy correlation matrix, which indicates that passenger occupancy also has strong local and long-range correlations. Similarly, passenger occupancy is also correlated with departure headway. In the third panel, we observe the cross-correlation matrix between link travel time and passenger occupancy, and we can see that they have strong correlations. Consequently, jointly modeling link travel time and passenger occupancy could make more accurate forecasting. The last two panels show the cross-correlation matrix between the leading bus and the following bus in terms of link travel time and passenger occupancy, respectively. We can see that two adjacent buses are strongly correlated in link travel time and passenger occupancy. Therefore, it is also crucial to take into account the interaction between two adjacent buses to improve prediction performance. In summary, our empirical findings demonstrate three key observations: i) strong interactions/correlations between travel time and passenger occupancy, ii) complex link correlations in travel time and passenger occupancy, and iii) strong interactions/correlations between adjacent buses along a bus route. Thus, we contend that, in constructing a forecasting model, bus travel time and occupancy should be jointly modeled, with explicit consideration given to the interactions between two adjacent buses. By doing so, the model will provide a more accurate representation of the dynamic and interconnected nature of bus travel time and passenger occupancy, thus improving the predictive performance of the model.

\section{Proposed model and Bayesian inference}\label{sec:model}

\subsection{Notations of variables}

\begin{figure}[!t]
\centering
\includegraphics[width = 0.75\textwidth]{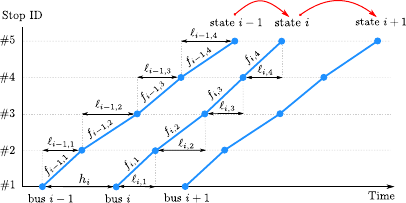}
\caption{Representation of bus trajectories with passenger occupancy.}
\label{fig:representation_pro}
\end{figure}

Let $\ell_{i,m}^{\left(d\right)}$ represent the travel time of the $i$-th bus on the $m$-th link on the $d$-th day. The trip travel time of the $i$-th bus from stop $m_1$ to stop $m_2$ is given by $\sum_{m=m_1}^{m_2 - 1} \ell_{i,m}^{\left(d\right)}$. We denote by $f_{i,m}^{\left(d\right)}$ the passenger occupancy of the $i$-th bus on the $m$-th link on the $d$-th day. The link travel time and occupancy of bus $i$ on a bus route with $n$ links (i.e., $n+1$ bus stops) can be stacked into $\boldsymbol{\ell}_i^{\left(d\right)}=\left[\ell_{i,1}^{\left(d\right)},\ell_{i,2}^{\left(d\right)},\cdots,\ell_{i,n}^{\left(d\right)}\right]^\top\in \mathbb{R}^n$ and $\boldsymbol{f}_i^{\left(d\right)}=\left[f_{i,1}^{\left(d\right)},f_{i,2}^{\left(d\right)},\cdots,f_{i,n}^{\left(d\right)}\right]^\top\in \mathbb{R}^n$, respectively. Note that here we define both travel time and occupancy to be real numbers for simplicity, whereas real-world travel time data should be strictly positive, and occupancy data should be in the form of counts. We define the departure headway $h_{i}^{\left(d\right)}$ as the time interval between the arrival of the $(i-1)$-th bus and the $i$-th bus at the originating bus stop on the $d$-th day. Fig.~\ref{fig:representation_pro} shows the representation of the variables in bus trajectories. We introduce a multivariate random variable, $\boldsymbol{y}_i^{\left(d\right)}$, as the concatenation of link travel time, the occupancy of the passengers and the departure progress for the bus $i$ on the $d$-th day:
\begin{equation}
\label{equ:fullx}   \boldsymbol{y}_{i}^{\left(d\right)}=\left[{\boldsymbol{\ell}_{i}^{\left(d\right)}}^\top, {\boldsymbol{f}_{i}^{\left(d\right)}}^\top,{h}_{i}^{\left(d\right)}\right]^\top\in\mathbb{R}^{2n+1}.
\end{equation}
In doing so, we create a concise representation that allows us to analyze and model the relationship among these variables more effectively. Therefore, from historical smart card data, we could collect the observations $\left\{\boldsymbol{y}_i^{\left(d\right)} \right\}_{i=1,d=1}^{I_d,D}$ denoted by $\mathcal{Y}$, where $I_d$ is the total number of bus runs we have in the training/historical data on the $d$-th day and $D$ represents the number of days in the data set.

\subsection{A Bayesian Markov regime-switching vector autoregressive model}

To model the random variable, \cite{chen2022probabilistic} developed a Bayesian time-dependent Gaussian mixture model for probabilistic forecasting. This model assumes that each pair of buses has a latent class (i.e., hidden state), and the core idea is to use the observed information to infer the hidden state. In Fig.~\ref{fig:representation_pro}, we can observe two successive bus pairs: the first pair $i$ (with bus $i-1$ and bus $i$) and the second pair $i+1$ (with bus $i$ and $i+1$). In the Bayesian Gaussian mixture model, the relationship between hidden states of adjacent bus pairs (e.g., pair $i$ and pair $i+1$) is not modeled directly. In other words, the state of bus pair $i+1$ has no direct relationship with the state of bus pair $i$ except that they likely come from the same time window. Consequently, when estimating the hidden state of a bus, the accuracy of the estimation heavily relies on the amount of observed information available for that bus. However, as mentioned in Section~\ref{sec:data}, the interactions between adjacent buses, particularly in scenarios where bus bunching occurs, reveal a clear interdependence between adjacent states. By accurately modeling the relationships between the states of adjacent buses, we can leverage more information to infer the state of the current bus and improve the estimation. For instance, in Fig.~\ref{fig:representation_pro}, when we want to estimate the hidden state of bus $i+1$, even if we have limited observed links for the current bus $i+1$, we can accurately estimate the state of bus $i$ because this bus could have more observed information. Using the modeled relationship between adjacent states and the observed links of bus $i+1$, we can enhance the estimation of the state of bus $i+1$. Therefore, considering the modeling of relationships between adjacent hidden states becomes crucial for accurate probabilistic forecasting. To address this issue, we propose employing a Bayesian regime-switching Markov model to capture the dependency between adjacent hidden states and to capture the relationship between two adjacent buses.

Markov regime-switching models are designed to analyze and forecast time series data that may shift between different states or regimes over time. These models have gained significant popularity in econometrics and time series analysis since the work of \cite{hamilton1989new}. Regime-switching models have been applied to many tasks such as speech recognition \citep{kim2017state} and motion recognition \citep{10.1214/14-AOAS742}.


In our Bayesian Markov regime-switching vector autoregressive model, we assume that the conditional distribution of the vector $\boldsymbol{y}_i^{\left(d\right)}$, given the preceding observed vector $\boldsymbol{y}_{i-1}^{\left(d\right)}$ and the current hidden state $z_i^{\left(d\right)}$, is described by
\begin{equation}
    \boldsymbol{y}_i^{\left(d\right)}\mid \boldsymbol{y}_{i-1}^{\left(d\right)},z_i^{\left(d\right)}=k \sim \mathcal{N}\left(\boldsymbol{A}_{k}\boldsymbol{y}_{i-1}^{\left(d\right)}+\boldsymbol{\mu}_{k},\boldsymbol{\Sigma}_{k}\right), \label{xpdf}
\end{equation}
where $\boldsymbol{y}_i^{\left(d\right)}\mid \boldsymbol{y}_{i-1}^{\left(d\right)},z_i^{\left(d\right)}=k$ follows a multivariate Gaussian distribution. This model is commonly used to capture complex and heterogeneous relationships among variables in multivariate time series analysis \citep{krolzig2013markov}. In this equation, $\boldsymbol{A}_k$ is a coefficient matrix that characterizes the influence of the preceding bus on the current bus under the regime of the hidden state $k$. The mean vector $\boldsymbol{\mu}_k$ and the covariance matrix $\boldsymbol{\Sigma}_k$ are parameters specific to the hidden state $k$, allowing the model to capture the varying dynamics of the bus system under different operational conditions. The hidden state $z_i^{\left(d\right)}$ represents the latent regime or condition of the $i$-th bus on the $d$-th day. These hidden states encapsulate unobserved factors that influence the bus's operational characteristics, such as traffic conditions, weather, or other variables. By modeling $\boldsymbol{y}_i^{\left(d\right)}$ as a function of both its preceding observation $\boldsymbol{y}_{i-1}^{\left(d\right)}$ and the hidden state $z_i^{\left(d\right)}$, the model can capture the complex and dynamic interactions within the bus route, thus improving the predictive accuracy of bus travel time and passenger occupancy over nonstationary operational conditions.  

Another critical aspect of our proposed model is the transition of hidden states, represented by probability $p\left(z_i^{\left(d\right)} \mid z_{i-1}^{\left(d\right)}\right)$. This probability is modeled as a categorical distribution dependent on the state transition matrix $\boldsymbol{\pi}$. We use $\boldsymbol{\pi}_k=\left(\pi_1,\ldots,\pi_K\right)^\top$ to denote a vector representing the transition probabilities from state $k$ to other states, i.e., $0\le\pi_k\le 1$ for $k=1,\ldots,K$ and $\sum_{k=1}^K \pi_k=1$, and $\boldsymbol{\pi} = \left[\boldsymbol{\pi}_1^\top,\dots,\boldsymbol{\pi}_K^\top\right]^\top$ represents the state transition matrix. Specifically, the probability of transition from state $z_{i-1}^{\left(d\right)}$ to state $z_i^{\left(d\right)}$ is given by $\pi_{z_{i-1}^{\left(d\right)},z_{i}^{\left(d\right)}}$, which follows a categorical distribution parameterized by $\boldsymbol{\pi}_{z_{i-1}^{\left(d\right)}}$
\begin{equation}
    z_i^{\left(d\right)} \mid z_{i-1}^{\left(d\right)} \sim \text{Categorical}\left(\boldsymbol{\pi}_{z_{i-1}^{\left(d\right)}}\right). \label{eq:zpdf}
\end{equation}
This formulation captures the Markov property of the model, where the probability of bus $i$ being in a particular state is dependent solely on the state of its preceding bus $i-1$. This structure is critical in modeling how the state of each bus is influenced by its immediate predecessor. For the initial state, we assume
\begin{equation}
    z_1^{\left(d\right)}\sim \text{Categorical}\left(\boldsymbol{\pi}^*\right),\label{sample_z1}
\end{equation}
where $\boldsymbol{\pi}^*$ is the probability distribution of the initial state, which is the marginal distribution calculated from the state transition matrix $\boldsymbol{\pi}$.

\subsection{Prior specification}
In general, we assume the following conjugate prior distributions for $\boldsymbol{\Sigma}_k$, $\boldsymbol{\mu}_k$, and $\boldsymbol{A}_k$,
\begin{align}
    \boldsymbol{\Sigma}_k & \sim \mathcal{W}^{-1}\left(\boldsymbol{\Psi}_0,\nu_0\right), \label{eq:cov prior} \\
    \boldsymbol{\mu}_k &\sim \mathcal{N}\left(\boldsymbol{\mu}_{0}, \frac{1}{\lambda_{0}}\boldsymbol{\Sigma}_k \right), \label{eq:mean prior} \\
    \boldsymbol{A}_k &\sim \mathcal{MN}\left(\boldsymbol{M}_{0},\boldsymbol{\Sigma}_k,\boldsymbol{V}_0 \right), \label{eq:A prior}
\end{align}
where $\mathcal{W}^{-1}\left(\boldsymbol{\Psi}_0,\nu_0\right)$ is the inverse-Wishart distribution with a scale matrix $\boldsymbol{\Psi}_0$ and $\nu_0$ degrees of freedom; $\boldsymbol{\mu}_0$ and $\lambda_0$ are parameters for the Gaussian prior; $\mathcal{MN}\left(\boldsymbol{M}_{0},\boldsymbol{\Sigma}_k,\boldsymbol{V}_0 \right)$ is the matrix Gaussian distribution with parameters $\boldsymbol{M}_{0}$, $\boldsymbol{\Sigma}_k$, and $\boldsymbol{V}_0$. These prior distributions are fundamental to our Bayesian framework, allowing the model to incorporate prior knowledge and uncertainty effectively.
In addition, we use a Dirichlet prior distribution for each transition probability:
\begin{equation}
    \boldsymbol{\pi}_k\mid \boldsymbol{\alpha} \sim \text{Dirichlet}\left(\boldsymbol{\alpha}\right),\label{eq:pi prior}
\end{equation}
where $\boldsymbol{\alpha}$ is the concentration parameter of the Dirichlet distribution. The Dirichlet distribution is a natural choice for modeling probability vectors $\boldsymbol{\pi}_k$ because it ensures that the probabilities are non-negative and sum up to one, which are essential properties for any set of transition probabilities.
\subsection{Model overview}
\begin{figure}[!t]
\centering
\includegraphics[width = 0.55\textwidth]{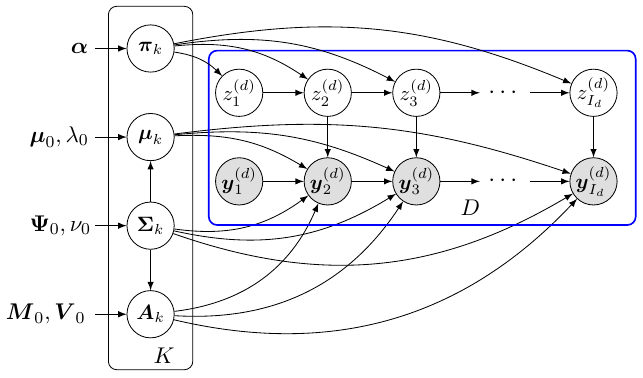}
\caption{Graphical representation of a Bayesian Markov regime-switching vector autoregressive model.}
\label{fig:graphical model}
\end{figure}

Fig.~\ref{fig:graphical model} provides the overall graphical representation of the proposed model. Considering the vector autoregressive model and the Markov regime-switching model as separate dimensions, we can think of the proposed combination as a spatial-temporal model. The overall model specification is summarized as follows:

\hspace{0pt} (1) Draw model parameters of state $k$ from 1 to $K$:

\hspace{20pt} (a) Draw the state transition probability $\boldsymbol{\pi}_k\mid \boldsymbol{\alpha} \sim \text{Dirichlet}\left(\boldsymbol{\alpha}\right)$.

\hspace{20pt} (b) Draw the covariance matrix $\boldsymbol{\Sigma}_k \sim \mathcal{W}^{-1}\left(\boldsymbol{\Psi}_0,\nu_0\right)$.

\hspace{20pt} (c) Draw the mean vector $\boldsymbol{\mu}_k \sim \mathcal{N}\left(\boldsymbol{\mu}_{0}, \frac{1}{\lambda_{0}}\boldsymbol{\Sigma}_k \right)$.

\hspace{20pt} (d) Draw the coefficient matrix $\boldsymbol{A}_k \sim \mathcal{MN}\left(\boldsymbol{M}_{0},\boldsymbol{\Sigma}_k,\boldsymbol{V}_0 \right)$.

\hspace{0pt} (2) For each sequence of $d$-th day from 1 to $D$:

\hspace{20pt} (a) Draw the initial state $z_1^{\left(d\right)}\sim \text{Categorical}\left(\boldsymbol{\pi}^*\right)$.

\hspace{20pt} (b)
For each bus $i$ from 2 to $I_d$:

\hspace{40pt} (i) Draw state sequence $z_i^{\left(d\right)}\mid z_{i-1}^{\left(d\right)}\sim \text{Categorical}\left(\boldsymbol{\pi}_{z_{i-1}^{\left(d\right)}}\right)$.

\hspace{40pt} (ii) Draw observations $\boldsymbol{y}_i^{\left(d\right)}\mid \boldsymbol{y}_{i-1}^{\left(d\right)},z_i^{\left(d\right)}=k \sim \mathcal{N}\left(\boldsymbol{A}_{k}\boldsymbol{y}_{i-1}^{\left(d\right)}+\boldsymbol{\mu}_{k},\boldsymbol{\Sigma}_{k}\right)$.


\subsection{MCMC algorithm to approximate posterior distribution}
\label{sec:MCMC}
For any $k=1,\ldots,K$, we define $M_k=\left\{(i,d)\mid z_i^{\left(d\right)}=k, d=1,\ldots,D, i=1,\ldots,I_d\right\}$ as the set of bus indices with state $z_i^{\left(d\right)}=k$; we use $\left|M_k\right|$ to denote the number of elements in $M_k$. Similarly, for any $k=1,\ldots,K$ and $k^\prime=1,\ldots,K$, we define $M_{k,k^\prime
}=\left\{(i,d) \mid z_i^{\left(d\right)}=k, z_{i+1}^{\left(d\right)}=k^\prime, d=1,\ldots,D,i=1,\ldots,I_d-1\right\}$ as the set of indices of buses that have $z_i^{\left(d\right)}=k$ with its follower having $z_{i+1}^{\left(d\right)}=k^\prime$. We define $\boldsymbol{A}=\left\{\boldsymbol{A}_k\mid k=1,\dots,K\right\}$ as the set of coefficient matrices. Define $\boldsymbol{\mu}=\left\{\boldsymbol{\mu}_k\mid k=1,\dots,K\right\}$ as the set of mean vectors and $\boldsymbol{\Sigma}=\left\{\boldsymbol{\Sigma}_k\mid k=1,\dots,K\right\}$ as the set of covariance matrices. For simplicity, we let $\Theta=\left\{\boldsymbol{\mu}_0,\lambda_0, \boldsymbol{\Psi}_0, \nu_0\right\}$ denote the set of hyperparameters for the Gaussian-inverse-Wishart prior distribution in Eq.~\eqref{eq:cov prior} and \eqref{eq:mean prior}. In addition, we let $\boldsymbol{y}^{\left(d\right)}_{1:I_d}$ and $\boldsymbol{z}^{\left(d\right)}_{1:I_d}$ denote the observations and states of the bus sequence on the $d$-th day; we denote by $\mathcal{Y}_{k}=\left\{\boldsymbol{y}_i^{\left(d\right)}\mid z_i^{\left(d\right)}=k, d=1,\ldots,D,i=1,\ldots,I_d\right\}$ the set of data vectors belonging to state $k$. Let $\boldsymbol{Z}=\left\{z_i^{\left(d\right)}\mid d=1,\ldots,D,i=1,\ldots,I_d\right\}$ to be the set of states of observations. We use $\boldsymbol{\Gamma} = \left\{\boldsymbol{A},\boldsymbol{\mu},\boldsymbol{\Sigma},\boldsymbol{Z}\right\}$ to denote a parameter set. The likelihood of $\boldsymbol{A}$, $\boldsymbol{\mu}$, $\boldsymbol{\Sigma}$, $\boldsymbol{Z}$ given observations $\mathcal{Y}$ is given by
\begin{align}
    L\left(\mathcal{Y},\boldsymbol{\Gamma}\right) = \prod_{d=1}^D\prod_{i=2}^{I_d}&p\left(\boldsymbol{y}_i^{\left(d\right)}\mid\boldsymbol{y}_{i-1}^{\left(d\right)},z_i^{\left(d\right)},\boldsymbol{A}_{z_{i}^{\left(d\right)}},\boldsymbol{\mu}_{z_{i}^{\left(d\right)}},\boldsymbol{\Sigma}_{z_{i}^{\left(d\right)}}\right)p\left(\boldsymbol{A}_{z_{i}^{\left(d\right)}}\mid\boldsymbol{M}_0,\boldsymbol{\Sigma}_{z_{i}^{\left(d\right)}},\boldsymbol{V}_0\right)p\left(\boldsymbol{\mu}_{z_{i}^{\left(d\right)}}\mid\boldsymbol{\mu}_0,\frac{1}{\lambda_0}\boldsymbol{\Sigma}_{z_{i}^{\left(d\right)}}\right) \notag\\
    &p\left(\boldsymbol{\Sigma}_{z_{i}^{\left(d\right)}}\mid\boldsymbol{\Phi}_0,\nu_0\right)\prod_{d=1}^Dp\left(z_1^{\left(d\right)}\mid\boldsymbol{\pi}^*\right)\prod_{i=2}^{I_d}p\left(z_{i+1}^{\left(d\right)}\mid z_i^{\left(d\right)},\boldsymbol{\pi}\right)p\left(\boldsymbol{\pi}\mid\boldsymbol{\alpha}\right).\label{eq:likelihood}
\end{align}
Due to the large number of observations, it is impossible to marginalize out $\boldsymbol{Z}$ from Eq.~\eqref{eq:likelihood}. Therefore, based on the graphical model illustrated in Fig.~\ref{fig:graphical model}, we derive an efficient MCMC scheme using Gibbs sampling. We start the Gibbs sampling with random initialization for
all parameters, and then iteratively sample each parameter from its conditional distribution on other
parameters. Posterior full conditional distributions of the parameters are derived as follows:
\begin{itemize}
    \item \textbf{Sampling state transition probability $\boldsymbol{\pi}_k$ from $p\left(\boldsymbol{\pi}_k \mid \boldsymbol{z}^{k}, \boldsymbol{\alpha}\right)$.} We use $\boldsymbol{z}^{k}$ to denote the vector containing latent variables whose values are $k$. The conditional distribution is $p\left(\boldsymbol{\pi}_k\mid \boldsymbol{z}^{k},\boldsymbol{\alpha}\right) \propto p\left(\boldsymbol{\pi}_k\mid \boldsymbol{\alpha}\right) p\left(\boldsymbol{z}^{k}\mid \boldsymbol{\pi}_k\right)$. The prior distribution $p\left(\boldsymbol{\pi}_k\mid \boldsymbol{\alpha}\right) = \text{Dirichlet}\left(\boldsymbol{\pi}_k\mid \boldsymbol{\alpha}\right) \propto \prod_{k^\prime=1}^{K}{\pi}_{k,k^\prime}^{\alpha_{k^\prime}-1}$, and $p\left(\boldsymbol{z}^{k}\mid \boldsymbol{\pi}_k\right)$ can be seen as a multinomial distribution $p\left(\boldsymbol{z}^{k}\mid \boldsymbol{\pi}_k\right)=\text{Multinomial}_K\left(\boldsymbol{z}^{k}\mid N,\boldsymbol{\pi}_k\right)\propto \prod_{k^\prime=1}^{K}{\pi}_{k,j}^{\left|M_{k,k^\prime}\right|}$, where $\left|M_{k,k^\prime}\right|$ is the number of elements in $M_{k,k^\prime}$. Therefore, the conditional posterior distribution is a Dirichlet distribution:
    \begin{equation}\label{sample_pi}
    p\left(\boldsymbol{\pi}_k\mid \boldsymbol{z}^{k},\boldsymbol{\alpha}\right)= \text{Dirichlet}\left(\left|M_{k,1}\right|+\alpha_1,\left|M_{k,2}\right|+\alpha_2,\cdots,\left|M_{k,K}\right|+\alpha_K\right).
    \end{equation}
    \item \textbf{Sampling state sequence $\boldsymbol{z}_{1:I_d}^{\left(d\right)}$ from $p\left(\boldsymbol{z}_{1:I_d}^{\left(d\right)} \mid \boldsymbol{y}_{1:I_d}^{\left(d\right)},\boldsymbol{\pi},\boldsymbol{\mu},\boldsymbol{\Sigma},\boldsymbol{A}\right)$.} For each bus sequence, we sample the entire hidden discrete state sequence $\boldsymbol{z}_{1:I_d}^{\left(d\right)}$ all at once given the sequence observations $\boldsymbol{y}_{1:I_d}^{\left(d\right)}$ and parameters $\boldsymbol{\pi},\boldsymbol{\mu},\boldsymbol{\Sigma},\boldsymbol{A}$. For the Markov chain model, the forward-backward algorithm is often used to sample the state sequence \citep{scott2002bayesian}. In this sampling algorithm, we are interested in finding the posterior distribution $p\left(z_i^{\left(d\right)}\mid\boldsymbol{y}_{1:I_d}^{\left(d\right)},\boldsymbol{\pi},\boldsymbol{\mu},\boldsymbol{\Sigma},\boldsymbol{A}\right)$ of $z_i^{\left(d\right)}$ given the sequence observations $\boldsymbol{y}_{1:I_d}^{\left(d\right)}$ and parameters $\boldsymbol{\pi},\boldsymbol{\mu},\boldsymbol{\Sigma},\boldsymbol{A}$. With Bayes' theorem, we have
    \begin{align}
        p\left(z_i^{\left(d\right)}\mid\boldsymbol{y}_{1:I_d}^{\left(d\right)},\boldsymbol{\pi},\boldsymbol{\mu},\boldsymbol{\Sigma},\boldsymbol{A}\right) = \frac{p\left(\boldsymbol{y}_{1:I_d}^{\left(d\right)}\mid z_i^{\left(d\right)},\boldsymbol{\mu},\boldsymbol{\Sigma},\boldsymbol{A}\right)p\left(z_i^{\left(d\right)}\mid \boldsymbol{\pi}\right)}{p\left(\boldsymbol{y}_{1:I_d}^{\left(d\right)}\mid\boldsymbol{\pi},\boldsymbol{\mu},\boldsymbol{\Sigma},\boldsymbol{A}\right)}.
    \end{align}
    Using the conditional independence property in Fig.~\ref{fig:graphical model}, we obtain
    \begin{align}
        p\left(z_i^{\left(d\right)}\mid\boldsymbol{y}_{1:I_d}^{\left(d\right)},\boldsymbol{\pi},\boldsymbol{\mu},\boldsymbol{\Sigma},\boldsymbol{A}\right) &= \frac{p\left(\boldsymbol{y}_{1:i}^{\left(d\right)},z_i^{\left(d\right)}\mid\boldsymbol{\pi},\boldsymbol{\mu},\boldsymbol{\Sigma},\boldsymbol{A}\right)p\left(\boldsymbol{y}_{i+1:I_d}^{\left(d\right)}\mid z_i^{\left(d\right)},\boldsymbol{\mu},\boldsymbol{\Sigma},\boldsymbol{A}\right)}{p\left(\boldsymbol{y}_{1:I_d}^{\left(d\right)}\mid\boldsymbol{\pi},\boldsymbol{\mu},\boldsymbol{\Sigma},\boldsymbol{A}\right)}\notag\\
        &= \frac{\alpha\left(z_i^{\left(d\right)}\right)\beta\left(z_i^{\left(d\right)}\right)}{p\left(\boldsymbol{y}_{1:I_d}^{\left(d\right)}\mid\boldsymbol{\pi},\boldsymbol{\mu},\boldsymbol{\Sigma},\boldsymbol{A}\right)}, \label{eq:fb}
    \end{align}
where we define $\alpha\left(z_i^{\left(d\right)}\right)$ and $\beta\left(z_i^{\left(d\right)}\right)$ as follows
\begin{align}
    \alpha\left(z_i^{\left(d\right)}\right) &= p\left(\boldsymbol{y}_{1:i}^{\left(d\right)},z_i^{\left(d\right)}\mid\boldsymbol{\pi},\boldsymbol{\mu},\boldsymbol{\Sigma},\boldsymbol{A}\right),\\
    \beta\left(z_i^{\left(d\right)}\right)&=p\left(\boldsymbol{y}_{i+1:I_d}^{\left(d\right)}\mid z_i^{\left(d\right)},\boldsymbol{\mu},\boldsymbol{\Sigma},\boldsymbol{A}\right).
\end{align}
We then derive recursive relationships that allow $\alpha\left(z_i^{\left(d\right)}\right)$ and $\beta\left(z_i^{\left(d\right)}\right)$ to be evaluated efficiently. The relationship between $\alpha\left(z_i^{\left(d\right)}\right)$ and $\alpha\left(z_{i-1}^{\left(d\right)}\right)$ can be derived as
\begin{align}
    \alpha\left(z_i^{\left(d\right)}\right) &= p\left(\boldsymbol{y}_{i}^{\left(d\right)}\mid \boldsymbol{y}_{i-1}^{\left(d\right)},z_i^{\left(d\right)},\boldsymbol{\mu},\boldsymbol{\Sigma},\boldsymbol{A}\right)\sum_{z_{i-1}^{\left(d\right)}}p\left(\boldsymbol{y}_{1:i-1}^{\left(d\right)}\mid z_{i-1}^{\left(d\right)},\boldsymbol{\mu},\boldsymbol{\Sigma},\boldsymbol{A}\right)p\left(z_i^{\left(d\right)}\mid z_{i-1}^{\left(d\right)}\right)p\left(z_{i-1}^{\left(d\right)}\mid\boldsymbol{\pi}\right)\notag\\
    &= p\left(\boldsymbol{y}_{i}^{\left(d\right)}\mid \boldsymbol{y}_{i-1}^{\left(d\right)},z_i^{\left(d\right)},\boldsymbol{\mu},\boldsymbol{\Sigma},\boldsymbol{A}\right)\sum_{z_{i-1}^{\left(d\right)}}\alpha\left(z_{i-1}^{\left(d\right)}\right)p\left(z_i^{\left(d\right)}\mid z_{i-1}^{\left(d\right)}\right)\notag\\
    &= \mathcal{N}\left(\boldsymbol{y}_{i}^{\left(d\right)}\mid\boldsymbol{A}_{z_i^{\left(d\right)}}\boldsymbol{y}_{i-1}^{\left(d\right)}+\boldsymbol{\mu}_{z_i^{\left(d\right)}},\boldsymbol{\Sigma}_{z_i^{\left(d\right)}}\right)\sum_{z_{i-1}^{\left(d\right)}}\alpha\left(z_{i-1}^{\left(d\right)}\right)\pi_{z_{i-1}^{\left(d\right)}}\left(z_i^{\left(d\right)}\right).
\end{align}
We can similarly derive the recursive relationship for the quantities $\beta\left(z_i^{\left(d\right)}\right)$ as follows
\begin{align}
    \beta\left(z_i^{\left(d\right)}\right) &= \sum_{z_{i+1}^{\left(d\right)}}p\left(\boldsymbol{y}_{i+2:I_d}^{\left(d\right)}\mid z_{i+1}^{\left(d\right)},\boldsymbol{\mu},\boldsymbol{\Sigma},\boldsymbol{A}\right)p\left(\boldsymbol{y}_{i+1}^{\left(d\right)}\mid \boldsymbol{y}_{i}^{\left(d\right)},z_{i+1}^{\left(d\right)},\boldsymbol{\mu},\boldsymbol{\Sigma},\boldsymbol{A}\right)p\left(z_{i+1}^{\left(d\right)}\mid z_{i}^{\left(d\right)}\right)\notag\\
    &= \sum_{z_{i+1}^{\left(d\right)}}\beta\left(z_{i+1}^{\left(d\right)}\right)p\left(\boldsymbol{y}_{i+1}^{\left(d\right)}\mid \boldsymbol{y}_{i}^{\left(d\right)},z_{i+1}^{\left(d\right)},\boldsymbol{\mu},\boldsymbol{\Sigma},\boldsymbol{A}\right)p\left(z_{i+1}^{\left(d\right)}\mid z_{i}^{\left(d\right)}\right)\notag\\
    &=
    \sum_{z_{i+1}^{\left(d\right)}}\beta\left(z_{i+1}^{\left(d\right)}\right)\mathcal{N}\left(\boldsymbol{y}_{i+1}^{\left(d\right)}\mid\boldsymbol{A}_{z_{i+1}^{\left(d\right)}}\boldsymbol{y}_{i}^{\left(d\right)}+\boldsymbol{\mu}_{z_{i+1}^{\left(d\right)}},\boldsymbol{\Sigma}_{z_{i+1}^{\left(d\right)}}\right)\pi_{z_{i}^{\left(d\right)}}\left(z_{i+1}^{\left(d\right)}\right).
\end{align}
Considering that the left-hand side in Eq.~\eqref{eq:fb} is a normalized distribution, the quantity $p\left(\boldsymbol{y}_{1:I_d}^{\left(d\right)}\mid\boldsymbol{\pi},\boldsymbol{\mu},\boldsymbol{\Sigma},\boldsymbol{A}\right)$ can be obtained as follows:
\begin{align}
    p\left(\boldsymbol{y}_{1:I_d}^{\left(d\right)}\mid\boldsymbol{\pi},\boldsymbol{\mu},\boldsymbol{\Sigma},\boldsymbol{A}\right) = \sum_{z_i^{\left(d\right)}}\alpha\left(z_i^{\left(d\right)}\right)\beta\left(z_i^{\left(d\right)}\right).
\end{align}
After running the recursion from $i=1,\ldots,I_t$ to obtain $\alpha\left(z_1^{\left(d\right)}\right),\ldots,\alpha\left(z_{I_d}^{\left(d\right)}\right)$ and the recursion from $i=I_t,\ldots,1$ to obtain $\beta\left(z_{1}^{\left(d\right)}\right),\ldots,\beta\left(z_{I_t}^{\left(d\right)}\right)$, then we could evaluate $p\left(z_i^{\left(d\right)}\mid\boldsymbol{y}_{1:I_d}^{\left(d\right)},\boldsymbol{\pi},\boldsymbol{\mu},\boldsymbol{\Sigma},\boldsymbol{A}\right)$ and sample the state sequence $\boldsymbol{z}_{1:I_{d}}^{\left(d\right)}$.

    \item \textbf{Sampling mean and covariance $(\boldsymbol{\mu}_k,\boldsymbol{\Sigma}_k)$ from $p\left(\boldsymbol{\mu}_k, \boldsymbol{\Sigma}_k \mid \mathcal{Y}_k, \Theta,\boldsymbol{A}_k\right)$.} Thanks to the conjugate prior distribution, the conditional distribution of the mean vector and the covariance matrix
$p\left(\boldsymbol{\mu}_k, \boldsymbol{\Sigma}_k \mid \mathcal{Y}_k, \Theta,\boldsymbol{A}_k \right)$ is a Gaussian-inverse-Wishart distribution:
\begin{equation}\label{mu123}
p\left(\boldsymbol{\mu}_k,\boldsymbol{\Sigma}_k \mid \mathcal{Y}_k, \Theta,\boldsymbol{A}_k \right) = \mathcal{N}\left(\boldsymbol{\mu}_k\mid\boldsymbol{\mu}_0^*,\frac{1}{\lambda_0^*}\boldsymbol{\Sigma}_k\right)\mathcal{W}^{-1}\left(\boldsymbol{\Sigma}_k\mid\boldsymbol{\Psi}_0^*,\nu_0^*\right),
\end{equation}
where
\begin{equation}\label{l123}
\begin{aligned}
&\boldsymbol{\mu}_0^* = \frac{\lambda_0\boldsymbol{\mu}_0+\left|M_k\right|\boldsymbol{\delta}}{\lambda_0+\left|M_k\right|},\quad \lambda_0^* = \lambda_0+\left|M_k\right|,\quad \nu_0^*=\nu_0+\left|M_k\right|, \\
& \boldsymbol{\delta}=\frac{1}{\left|M_k\right|}\sum_{i,d\in M_k}\left(\boldsymbol{y}_{i}^{\left(d\right)}-\boldsymbol{A}_k\boldsymbol{y}_{i-1}^{\left(d\right)}\right),\\
&
\boldsymbol{\Psi}_0^* = \boldsymbol{\Psi}_0+\boldsymbol{\mathrm{S}}+\frac{\lambda_0\left|M_k\right|}{\lambda_0+\left|M_k\right|}\left(\boldsymbol{\delta}-\boldsymbol{\mu_0}\right)\left(\boldsymbol{\delta}-\boldsymbol{\mu_0}\right)^{\top},  \\
& \boldsymbol{\mathrm{S}}=\sum_{i,d\in M_k}\left(\boldsymbol{y}_{i}^{\left(d\right)}-\boldsymbol{A}_k\boldsymbol{y}_{i-1}^{\left(d\right)}-\boldsymbol{\delta}\right)\left(\boldsymbol{y}_{i}^{\left(d\right)}-\boldsymbol{A}_k\boldsymbol{y}_{i-1}^{\left(d\right)}-\boldsymbol{\delta}\right)^{\top}.
\end{aligned}
\end{equation}
\item \textbf{Sampling coefficient matrix $\boldsymbol{A}_k$ from $p\left(\boldsymbol{A}_k \mid \mathcal{Y}_k, \boldsymbol{\mu}_k,\boldsymbol{\Sigma}_k\right)$.} As we use the conjugate prior distribution, the conditional distribution of the coefficient matrix
$p\left(\boldsymbol{A}_k \mid \mathcal{Y}_k, \boldsymbol{\mu}_k,\boldsymbol{\Sigma}_k \right)$ is a matrix Gaussian distribution:
\begin{equation}\label{A}
p\left(\boldsymbol{A}_k \mid \mathcal{Y}_k, \boldsymbol{\mu}_k,\boldsymbol{\Sigma}_k \right) = \mathcal{MN}\left(\boldsymbol{A}_k\mid\boldsymbol{M}_0^*,\boldsymbol{\Sigma}_k,\boldsymbol{V}_0^*\right),
\end{equation}
where
\begin{align}
    \boldsymbol{V}_0^* &= \left(\boldsymbol{V}_0^{-1}+\sum_{i,d\in M_k}\boldsymbol{y}_{i-1}^{\left(d\right)}{\boldsymbol{y}_{i-1}^{\left(d\right)}}^\top\right)^{-1},\\
    \boldsymbol{M}_0^* &= \left(\boldsymbol{M}_0\boldsymbol{V}_0^{-1}+\sum_{i,d\in M_k}\left(\boldsymbol{y}_i^{\left(d\right)}-\boldsymbol{\mu}_k\right){\boldsymbol{y}_{i-1}^{\left(d\right)}}^\top\right)\boldsymbol{V}_0^*.
\end{align}
\end{itemize}

Finally, we summarize the Gibbs sampling procedure to estimate the parameters in Algorithm~\ref{alg:gibbs} seen in \hyperref[appA]{Appendix A}. We drop the first $n_1=9000$ iterations as burn-in and then store samples of parameters $\boldsymbol{\pi}_k$, $\boldsymbol{\mu}_k$, $\boldsymbol{\Sigma}_k$, $\boldsymbol{A}_k$ from the following $n_2=1000$ iterations. 
For hyperparameters, we set $\boldsymbol{\mu}_0 = \boldsymbol{0}_{2n+1}$, $\lambda_0=2$,  $\boldsymbol{\Phi}_0 = \boldsymbol{I}_{2n+1}$,
$\boldsymbol{V}_0 = \boldsymbol{I}_{2n+1}$, $\nu_0 = 2n+3$, and $\boldsymbol{\alpha} = 0.2\times\boldsymbol{I}_{K}$, where $n$ is the number of bus links. Note that model training is in fact offline based on historical data, and only Markov chains (i.e. samples) of the parameters are used in the forecasting task. We code the MCMC algorithm using Python with Numpy and Scipy packages.

\section{Probabilistic forecasting}\label{sec:forecast} We categorize the links of each bus into two groups: observed links and upcoming links, during the forecasting process. Observed links refer to the links that the bus has already traversed, and we know their respective travel times and passenger occupancies. On the other hand, upcoming links are the links that the bus is yet to traverse, and we need to forecast their travel times and passenger occupancies. The crucial step of the forecasting process is to determine the hidden states of the buses using the available observed information. Once hidden states are identified, it becomes convenient to make probabilistic forecasts for travel time and occupancy for downstream links conditional on observed information. Through the model training process, we have obtained a sample from the posterior distribution of the parameters. Using sampling techniques, we can make estimations about the hidden states of buses using the information that is currently observed. Forecasting can be achieved by approximating the distribution of unobserved/future variables conditional on the observed link travel time and passenger occupancy. Typically, there are upcoming links for both the following bus (bus $j$) and the leading bus (bus $(j-1)$). While it is possible to make forecasts solely based on the observed links, we adopt an autoregressive approach. This approach incorporates the forecasting of the upcoming links of the leading bus to forecast the travel time and passenger occupancy of the following bus. When considering the interdependence among buses, our forecasting method offers a more comprehensive and accurate estimation.

\begin{figure}[!t]
\centering
\includegraphics[width = 0.78\textwidth]{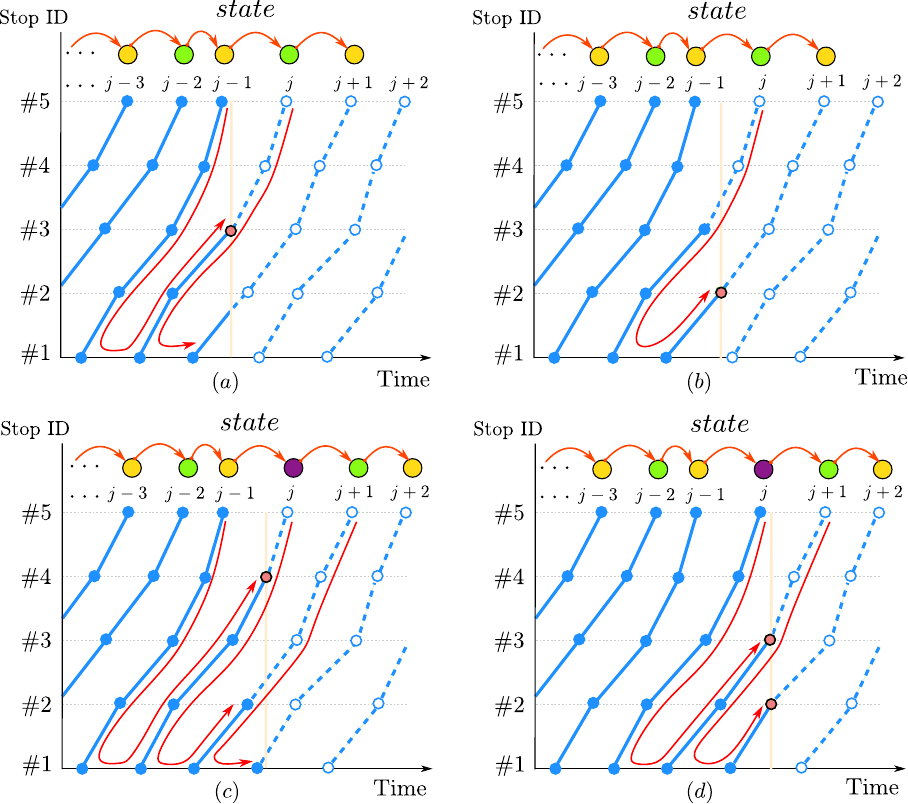}
\caption{Representation of probabilistic forecasting. Solid blue circles: observed link travel time and passenger occupancy at bus stops. The hollow circles: unknown arrival time and passenger occupancy at the bus stop. Solid red circles: new observed arrival time and passenger occupancy at bus stops. Solid green, orange, and purple circles represent different hidden states. Each sub-figure is a new round of forecasting triggered by a new observation of bus arrival. For each bus, we forecast the bus's upcoming link travel times and passenger occupancies (dashed blue lines) based on observed and previously forecast link travel times and passenger occupancies (blue lines covered by a red arrow area).}
\label{fig:rep4cast}
\end{figure}

We use Fig.~\ref{fig:rep4cast} to illustrate the forecasting process. Fig.~\ref{fig:rep4cast} (a) presents the scenario where bus $j-1$ has completed its run, bus $j$ has traversed the initial two links and arrived at stop $\#3$, and bus $j+1$ has left the origin stop but has not yet arrived at stop $\#2$. Our initial task is to estimate the hidden states of these buses (from bus 1 to bus $j+1$) based on the observed link travel times and passenger occupancies. Similarly, we used observed link information and headway to forecast the subsequent links for bus $j$ and bus $j+1$ based on the estimated states.  For bus $j$, we can make predictions by using its upstream links (the first two links), all the observed link travel times/passenger occupancies of the leading bus $j-1$, and the starting headway of the bus $j$. Concerning bus $j+1$, the observed upstream link travel times/passenger occupancies of bus $j$ are used, along with the forecasts of the upcoming link travel times/passenger occupancies of bus $j$, and the observed headway. As time passes, the forecasting of relevant buses can be updated upon receiving new observed links. At the time point illustrated in Fig.~\ref{fig:rep4cast} (b), bus $j+1$ obtains a new observed link, enabling us to update the hidden states and perform prediction. In this example, the states of all buses remain unchanged, but with additional observed links, the forecasting could become more accurate. At the subsequent time point depicted in Fig.~\ref{fig:rep4cast} (c), bus $j$ acquires a new observed link, which requires an update of the bus states. With more observed information, it is possible for the states of certain buses to change, resulting in improved accuracy. In this case, the state of bus $j$ changes from green to purple, and consequently we can update the forecast. It should be noted that even though bus $j+1$ does not have any additional observed links, updated information about its leading bus (i.e., bus $j$) can reinforce/enhance its forecasting. Furthermore, we can see that a new bus $j+2$ starts its trip on the route; with the observed headway, we could make forecasting for bus $j+2$. Fig.~\ref{fig:rep4cast} (d) demonstrates the scenario where bus $j$ has completed its run, and bus $j+1$ and $j+2$ each obtain a new observed link. In this case, we can generate forecasts for bus $j+1$ and $j+2$ following the aforementioned procedure.

Consider a specific bus $j$, and the observed links (travel time, passenger occupancy, headway) can be organized into a vector denoted as $\boldsymbol{y}_{j}^o$, representing partial observations. Our goal is to forecast the unobserved links of bus $j$, and we denote the vector we want to forecast as $\boldsymbol{y}_{j}^f$. Therefore, the full variable of bus $j$ is $\boldsymbol{y}_j = \left[{\boldsymbol{y}_{j}^o}^\top,{\boldsymbol{y}_{j}^f}^\top\right]^\top$. For most time-series problems, the Markov regime vector autoregressive model often makes forecasting based only on previous observations. For example, forecasts of $\boldsymbol{y}_j^f$ are derived solely from its immediate predecessor $\boldsymbol{y}_{j-1}$. This conventional approach relies on the dynamics inherent in the vector autoregressive structure and the probabilistic transitions between the hidden states. At any time, in additional to the observation $\boldsymbol{y}_{j-1}$, we also have partial observations $\boldsymbol{y}_j^o$ and $\boldsymbol{y}_{j+1}^o$. Our proposed forecasting approach will incorporate all observations (that is, $\boldsymbol{y}_{j-1}$, $\boldsymbol{y}_j^o$, $\boldsymbol{y}_{j+1}^o$) to enhance the forecast for the unobserved part $\boldsymbol{y}_j^f$ of bus $j$. With the collected $n_2$ samples $\left\{\boldsymbol{\theta}^{\left(\rho\right)}\right\}_{\rho=1}^{n_2}$ of parameters during the model estimation stage, we can utilize the Gibbs sampling method to obtain the predicted distribution (probabilistic forecasting) for unobserved variables. The predictive distribution over the unobserved part $\boldsymbol{y}_j^f$ of bus $j$ given the observed data can be approximated by Monte Carlo estimation:
\begin{align}
    &p\left(\boldsymbol{y}_j^f \mid \boldsymbol{y}_{1:j-1},\boldsymbol{y}_j^o,\boldsymbol{y}_{j+1}^o\right)\notag \\ &=\iiiint p\left(\boldsymbol{y}_j^f\mid \boldsymbol{y}_j^o,\boldsymbol{y}_{j-1},\boldsymbol{y}_{j+1}^o,z_j,z_{j+1},\boldsymbol{\theta}\right)p\left(z_{j},z_{j+1}\mid \boldsymbol{y}_j^o,\boldsymbol{y}_{1:j-1},\boldsymbol{y}_{j+1}^o,\boldsymbol{\pi},\boldsymbol{\theta}\right)p\left(\boldsymbol{\pi}\right)p\left(\boldsymbol{\theta}\right)dz_jdz_{j+1}d\boldsymbol{\pi}d\boldsymbol{\theta} \notag\\
    &\approx \frac{1}{n_2}\sum_{\rho=1}^{n_2}p\left(\boldsymbol{y}_j^f\mid\boldsymbol{y}_j^o,\boldsymbol{y}_{j-1},\boldsymbol{y}_{j+1}^o,z_j^{\left(\rho\right)},z_{j+1}^{\left(\rho\right)},\boldsymbol{\mu}^{\left(\rho\right)},\boldsymbol{\Sigma}^{\left(\rho\right)},\boldsymbol{A}^{\left(\rho\right)}\right).
\end{align}

\begin{figure}[!t]
\centering
\includegraphics[width = 0.5\textwidth]{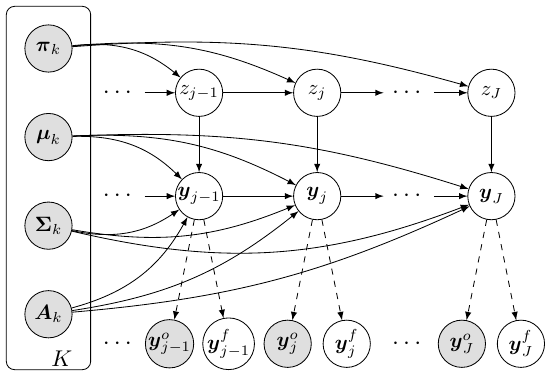}
\caption{Graphical representation of probabilistic forecasting. Each bus has the partial observations represented by a vector $\boldsymbol{y}_{j}^o$. Our goal is to forecast the unobserved links of the buses, and the vector we want to forecast is represented by $\boldsymbol{y}_{j}^f$.}
\label{fig:graphical4cast}
\end{figure}

Fig.~\ref{fig:graphical4cast} shows the overall graphical representation of probabilistic forecasting. The sampling scheme for probabilistic forecasting is as follows.

For the sequence state inference, we use the same forward-backward sampling method as in Section~\ref{sec:MCMC}. Next, we need to sample $\boldsymbol{y}_j^f$ from $p\left(\boldsymbol{y}_j^f\mid\boldsymbol{y}_j^o,\boldsymbol{y}_{j-1},\boldsymbol{y}_{j+1}^o,z_j,z_{j+1},\boldsymbol{\mu},\boldsymbol{\Sigma},\boldsymbol{A}\right)$. Note that in sampling $\boldsymbol{y}_j^f$ we build the conditional distribution on $\boldsymbol{y}_{j+1}^o$ instead of $\boldsymbol{y}_{j+1}$ for efficient sampling. Here, by using the characteristics of Gaussian linear systems, we can easily obtain the joint distribution of $\boldsymbol{y}_j$ and $\boldsymbol{y}_{j+1}$ as
\begin{equation}
p\left(\left[\begin{array}{c}
\boldsymbol{y}_{j}\\
\boldsymbol{y}_{j+1}
\end{array}\right] \middle| \boldsymbol{y}_{j-1}, z_j,z_{j+1},\boldsymbol{\mu},\boldsymbol{\Sigma},\boldsymbol{A} \right)\sim \mathcal{N}\left(\boldsymbol{m},\boldsymbol{L}\right),
\end{equation}
where
\begin{equation}
\boldsymbol{m} = \begin{bmatrix}
\boldsymbol{A}_{z_j}\boldsymbol{y}_{j-1}+\boldsymbol{\mu}_{z_j}\\
\boldsymbol{A}_{z_{j+1}}(\boldsymbol{A}_{z_j}\boldsymbol{y}_{j-1}+\boldsymbol{\mu}_{z_j})+\boldsymbol{\mu}_{z_{j+1}}\\
\end{bmatrix}, \boldsymbol{L} = \begin{bmatrix}
\boldsymbol{\Sigma}_{z_j} & {\boldsymbol{\Sigma}_{z_j}\boldsymbol{A}_{z_{j+1}}^\top} \\
{\boldsymbol{A}_{z_{j+1}}\boldsymbol{\Sigma}_{z_j}\quad} & \boldsymbol{A}_{z_{j+1}}\boldsymbol{\Sigma}_{z_j}\boldsymbol{A}_{z_{j+1}}^\top+\boldsymbol{\Sigma}_{z_{j+1}} \\
\end{bmatrix}.
\end{equation}

In this joint distribution, we have partial observations $\boldsymbol{y}^o=\left[{\boldsymbol{y}_j^o}^\top,{\boldsymbol{y}_{j+1}^o}^\top\right]^\top$. Based on the joint distribution, we can directly derive the conditional distribution of unobserved vectors $\boldsymbol{y}_j^f$:
\begin{equation}
p\begin{pmatrix}
\boldsymbol{y}_{j}^f\\\boldsymbol{y}^o
\end{pmatrix}\sim \mathcal{N} \left(\begin{bmatrix}
\boldsymbol{m}_{f}\\
\boldsymbol{m}_{o}\\
\end{bmatrix}, \begin{bmatrix}
\boldsymbol{L}_{f,f} & {\boldsymbol{L}_{f,o}} \\
{\boldsymbol{L}_{o,f}} & \boldsymbol{L}_{o,o} \\
\end{bmatrix}\right).
\end{equation}
The conditional distribution of $\boldsymbol{y}_j^f$ given $\boldsymbol{y}^o$ is:
\begin{align}
p\left(\boldsymbol{y}_j^f\mid\boldsymbol{y}_j^o,\boldsymbol{y}_{j-1},\boldsymbol{y}_{j+1}^o,z_j,z_{j+1},\boldsymbol{\mu},\boldsymbol{\Sigma},\boldsymbol{A}\right)=p\left(\boldsymbol{y}_j^f \mid \boldsymbol{y}^o\right) =  \mathcal{N}\left(\boldsymbol{y}_j^f\mid\boldsymbol{m}_{f\mid o},\boldsymbol{L}_{f\mid o}\right),\label{eq:conditional}
\end{align}
where $\boldsymbol{m}_{f\mid o}$ and $\boldsymbol{L}_{f\mid o}$ are the conditional mean and covariance matrix, respectively, given by:
\begin{align} \boldsymbol{m}_{f\mid o} &=  \boldsymbol{m}_f + \boldsymbol{L}_{f,o}\boldsymbol{L}_{o,o}^{-1}\left(\boldsymbol{y}^o-\boldsymbol{m}_{o}\right),
\\
\boldsymbol{L}_{f\mid o} &= \boldsymbol{L}_{f,f}-\boldsymbol{L}_{f,o}\boldsymbol{L}_{o,o}^{-1}\boldsymbol{L}_{o,f}.
\end{align}

By collecting forecasting samples from all parameter groups, we can approximate posterior predictive distributions for travel times and passenger occupancies of upcoming links. We summarize the Gibbs sampling algorithm for probabilistic forecasting in \hyperref[appB]{Appendix B}.

\section{Experiments}\label{sec:exp} In this section, we conduct a comprehensive evaluation of our proposed model using real-world datasets. We also undertake a comparative analysis with existing models to highlight the superior performance of our approach. In addition, we explore an examination of parameter patterns to further substantiate our findings. The source code used for these experiments can be accessed from \href{https://github.com/xiaoxuchen/Markov-Regime-switching-Model}{https://github.com/xiaoxuchen/Markov-Regime-switching-Model}.

\subsection{Experiment settings}
As the measurements have different units, we first perform data standardization (z-score normalization) so that all variables are centered at 0 with a standard deviation of 1. By doing so, we can better model and learn the covariance matrix. For example, $\ell_{i,m}$ (the $m$-th link travel time of the $i$-th bus) can be rescaled/standardized with
\begin{equation}
    \Tilde{\ell}_{i,m} =  \frac{\ell_{i,m}-\mu_{\ell_m}}{\sigma_{\ell_m}},
\end{equation}
where $\mu_{\ell_m}$ is the mean of travel time at the $m$-th link; $\sigma_{\ell_m}$ is the standard deviation of travel time at the $m$-th link. The dataset encompasses a period of four consecutive weeks, specifically focusing on weekdays (Monday to Friday), which amounts to a total of 20 days. We use the first 15 days for model inference, and the remaining 5 days to validate the model forecasting. In the inference, we can estimate the parameters of the model and understand the underlying structure or process that generates the observed data. The forecasting validation is crucial for evaluating how well our proposed model performs on unseen data. It helps in assessing the model's predictive accuracy and generalizability. In this experiment, we make probabilistic forecasting for bus travel time and passenger occupancy to evaluate the proposed model.


\subsection{Performance metrics}
We use the root mean squared error (RMSE) and the mean absolute error (MAE) to evaluate the performance for point estimation based on the mean. We use the continuous rank probability score (CRPS) to evaluate the performance of probabilistic forecasting.

\begin{itemize}
    \item  RMSE and MAE are defined as:
    \begin{equation}
    \begin{aligned}
         \text{RMSE}&=\sqrt{\frac{1}{n}\sum_{i=1}^{n}(y_i-\hat{y}_{i})^2}, \\
         \text{MAE}&=\frac{1}{n}\sum_{i=1}^{n}\left|{y_i-\hat{y}_i}\right|,
    \end{aligned}
    \end{equation}
    where $y_i,\hat{y}_{i},i=1,\ldots,n$ are the true values and forecasts, respectively.
    \item The continuous rank probability score is often used as a quantitative measure of probabilistic forecasting; it is defined as the quadratic measure of the discrepancy between the predicted cumulative distribution function (CDF, denoted by $F_{X}$) and $\mathbb{I}(x\geq y)$, the empirical CDF of the observation $y$:
    \begin{equation}
        \text{CRPS}(F_{X},y)=\int_{-\infty}^{\infty}\left(F_{X}(x)-\mathbb{I}(x\geq y)\right)^2dx,
    \end{equation}
    where $\mathbb{I}(\cdot)$ is the indicator function, which is defined as follows: If the condition inside the parentheses is true, then $\mathbb{I}(\cdot)$ equals 1. We use the average CRPS of all observations as one metric. Essentially, the CRPS calculates the mean squared difference between the predicted probabilities and the observed outcome, integrated over all possible threshold values. Lower CRPS values indicate better forecast accuracy.
\end{itemize}

\subsection{Model comparison} 
\begin{figure}[!t]
\centering
\includegraphics[width = 0.5\textwidth]{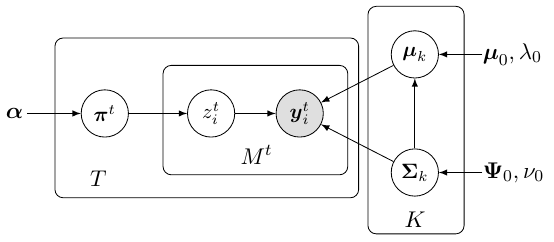}
\caption{Graphical representation of Bayesian time-dependent Gaussian mixture model.}
\label{fig:BGMM}
\end{figure}

In this paper, we compare the performance of our proposed model with the Bayesian time-dependent Gaussian mixture model developed in \cite{chen2022probabilistic}. Here, we detail the method outlined in Section~\ref{sec:intro}. Fig.~\ref{fig:BGMM} shows the overall graphical representation of the time-dependent Bayesian Gaussian mixture model. The random variable at the $t$-th period follows a multivariate Gaussian mixture model:
\begin{equation}\label{Multi-variate Density}
    p^{t}\left(\boldsymbol{y}^t \right) = \sum_{k=1}^{K}\pi_k^{t} \mathcal{N}\left(\boldsymbol{y}^t\mid\boldsymbol{\mu}_k,\boldsymbol{\Sigma}_k\right),
\end{equation}
where the superscript $(\cdot)^{t}$ denotes the time period, $K$ is the number of components, $0\le\pi_k^t\le 1$ is a mixing coefficient with $\sum_{k=1}^{K}\pi_k^t=1$, and each of the $K$ components follows a multivariate Gaussian distribution.

The random variable of each period is characterized by a mixture of several shared Gaussian distributions. In the graphical model, $z_i^t$ is a component label, indicating which component $\boldsymbol{y}_i^t$ belongs to. In a Bayesian setting, they use a conjugate Gaussian-inverse-Wishart prior on $\boldsymbol{\mu}_k$ and $\boldsymbol{\Sigma}_k$ and a Dirichlet prior on $\boldsymbol{\pi}^t$ for efficient inference. The overall data generation process is summarized as:
\begin{align}
    \boldsymbol{\pi}^t &\sim \text{Dirichlet}\left(\boldsymbol{\alpha}\right),
    \\
    \boldsymbol{\Sigma}_k& \sim \mathcal{W}^{-1}\left(\boldsymbol{\Psi}_0,\nu_0\right),  \\
    \boldsymbol{\mu}_k &\sim \mathcal{N}\left(\boldsymbol{\mu}_{0}, \frac{1}{\lambda_{0}}\boldsymbol{\Sigma}_k \right),  \\
    z_{i}^t &\sim \text{Categorical}\left(\boldsymbol{\pi}^t\right),
    \\
    \boldsymbol{y}_{i}^t\mid z_{i}^t =k&\sim \mathcal{N}\left(\boldsymbol{\mu}_{k},\boldsymbol{\Sigma}_{k}\right),
\end{align}
where $\boldsymbol{\alpha}$ is the concentration parameter of the Dirichlet distribution; $\mathcal{W}^{-1}\left(\boldsymbol{\Psi}_0,\nu_0\right)$ is the inverse-Wishart distribution with a scale matrix $\boldsymbol{\Psi}_0$ and $\nu_0$ degrees of freedom; $\boldsymbol{\mu}_0$ and $\lambda_0$ are parameters for the Gaussian prior.

Next, we design the following models to demonstrate the effect of interactions between bus travel time and passenger occupancy and dependencies between hidden states on probabilistic forecasting for bus travel time and passenger occupancy.

\begin{itemize}
    \item BGMM-S: Apply Bayesian time-dependent Gaussian mixture model to make bus travel time and passenger occupancy forecasting separately. We develop two independent models: one model for bus travel time and the other model for passenger occupancy. These independent models are the particular cases where the correlation between travel time and passenger occupancy is not considered.
    \item BGMM-J: Apply a Bayesian time-dependent Gaussian mixture model to make bus travel time and passenger occupancy forecasting jointly. We model the bus travel time and passenger occupancy as a random variable and therefore could consider the interactions between them.
    \item MSAR-S: Apply Bayesian Markov regime-switching vector autoregressive model to make bus travel time and passenger occupancy forecasting separately. Similarly, there are two independent models: one model for bus travel time and the other model for passenger occupancy.
    \item MSAR-J: This is our proposed model, and we utilize the Bayesian Markov regime-switching vector autoregressive model to jointly forecast bus travel time and passenger occupancy.
\end{itemize}

\subsection{Forecast performance}
We apply Algorithm~\ref{alg:gibbs} to estimate model parameters and Algorithm~\ref{alg:4cast} to make probabilistic forecasting for bus travel time and passenger occupancy. We test models with different numbers of clusters/states ($K=1,5,10,20,30,40,50$) to select an optimal state number. For each model, we start with an initial value $K=1$ and evaluate performance. When performance improves, we will select a larger $K$ to evaluate the model again and continue this process until there is no substantial improvement in performance. Table~\ref{tab: link} shows the performance of probabilistic forecasting for bus travel time and passenger occupancy with different models. First, we can see that all models show improved performance as the number of clusters/states ($K$) increases, indicating that they are significant in forecasting bus travel time and passenger occupancy. Second, we can observe that RSMM-J and BGMM-J outperform MSAR-S and MSAR-S, which demonstrates the importance of joint modeling of bus travel time and passenger occupancy. This finding shows that considering the interactions between bus travel time and passenger occupancy can significantly improve probabilistic forecast performance for bus travel time and passenger occupancy. Third, we can see that MSAR-S/J outperforms BGMM-S/J, which indicates that modeling states transition/connection could help make better probabilistic forecasting for bus travel time and passenger occupancy.

\begin{table}[!t]\centering
\caption{Performance of probabilistic forecasting of link travel time, passenger occupancy, and trip travel time.}\label{tab: link}
\scriptsize
\begin{tabular}{l|c|c|c|c|c|c|c|c|c|c}\toprule
\multicolumn{2}{c}{} &\multicolumn{3}{c|}{Link travel time (sec)} &\multicolumn{3}{c|}{Passenger occupancy (pax)}
&\multicolumn{3}{c}{Trip travel time (sec)}\\
\cmidrule{3-11}
\multicolumn{2}{c}{} &RMSE &MAE &CRPS &RMSE &MAE &CRPS &RMSE &MAE &CRPS \\
\midrule
\multirow{6}{*}{BGMM-S} &  $K = 1$ &64.31 &51.32 &32.44&10.05 &8.13 &6.20&258.83&215.61&175.18  \\
&$K = 5$ &54.51 &42.67 &29.60&9.16 &7.34 &5.87 &246.89&201.36&157.85\\
&$K = 10$ &47.41 &36.82 &27.44 &8.00 &6.89 &5.21&236.14&189.39&144.83 \\
&$K = 20$ &45.73 &35.54 &25.91 &7.81 &6.70 &5.43&218.65&176.94&132.44 \\
&$K = 30$ &\textbf{40.53} &\textbf{31.49} &\textbf{20.59} &7.67 &6.60 &5.37&\textbf{199.49}&\textbf{161.29}&\textbf{115.78} \\
&$K = 40$ &43.96 &33.97 &22.28 &\textbf{6.94} &\textbf{5.95} &\textbf{4.95}&212.09&172.35&125.93\\
\midrule
\multirow{7}{*}{BGMM-J} &$K = 1$ &47.23 &35.71 &27.25&8.97 &7.01 &5.45&221.09&184.14&135.29  \\
&$K = 5$ &44.63 &34.35 &25.69&8.64 &6.92 &5.28&213.53&174.15&131.21 \\
&$K = 10$ &38.94 &30.28 &19.96 &6.46 &5.27 &4.26&183.32&149.51&106.41 \\
&$K = 20$ &36.87 &28.32 &19.65 &6.16 &5.04 &3.96&180.88&143.22&103.60 \\
&$K = 30$ &24.16 &17.28 &17.63 &5.76 &4.69 &3.84&170.43&112.47&79.89 \\
&$K = 40$ &\textbf{18.25} &\textbf{13.02} &\textbf{14.35} &\textbf{4.53} &\textbf{3.60} &\textbf{3.57}&\textbf{164.22}&\textbf{102.23}&\textbf{72.36} \\
&$K = 50$ &18.41 &13.06 &14.64 &4.60 &3.74 &3.75 &166.80&105.52&73.99\\
\midrule
\multirow{5}{*}{MSAR-S} &$K = 1$ &62.87 &49.48 &31.15&10.23 &8.53 &6.32&257.11&214.04&174.87   \\
&$K = 5$ &51.86 &40.46 &25.32&7.65 &6.46&5.25&235.50&194.67&143.99 \\
&$K = 10$ &42.61 &32.73 &21.74 &6.89 &5.87 &4.73&205.99&169.55&120.36 \\
&$K = 20$ &\textbf{34.45} &\textbf{25.96} &\textbf{19.59} &\textbf{6.35} &\textbf{5.39} &\textbf{4.27}&\textbf{195.90}&\textbf{156.32}&\textbf{109.32} \\
&$K = 30$ &39.09 &30.28&20.01 &6.58 &5.61 &4.53&205.60&157.92&112.60 \\
\midrule
\multirow{6}{*}{MSAR-J} &$K = 1$ &48.32 &35.99 &27.45&9.20 &7.03 &5.48&222.28&184.37&135.60   \\
&$K = 5$ &39.38 &30.36 &20.39&6.43 &5.45 &4.82&197.60&157.43&114.00 \\
&$K = 10$ &30.16 &22.57 &18.25 &5.27 &4.41 &3.90&190.47&141.01&105.90 \\
&$K = 20$ &18.35 &13.38 &14.36 &4.86 &4.07 &3.79&164.47&103.55&72.60 \\
&$K = 30$ &\textbf{16.11} &\textbf{11.66} &\textbf{12.14} &\textbf{3.48} &\textbf{2.92} &\textbf{3.07}&\textbf{137.13}&\textbf{83.48}&\textbf{57.98} \\
&$K = 40$ &17.02 &12.57 &13.37 &4.50 &3.98 &3.71&153.35&95.72&63.06 \\
\bottomrule
\multicolumn{8}{l}{{Best results are highlighted in bold fonts.}}
\end{tabular}
\end{table}

\subsection{Interpreting analysis} 
\begin{figure}[!ht]
\centering
\includegraphics[width = 0.28\textwidth]{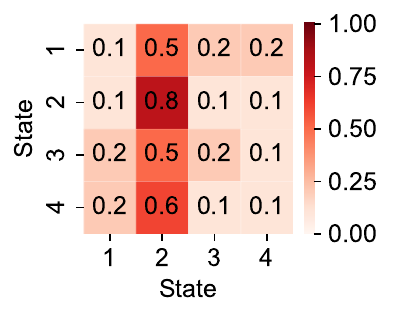}
\caption{Estimated transition matrix.}
\label{fig:transition}
\end{figure}

\begin{figure}[!ht]
\centering
\includegraphics[width = 0.8\textwidth]{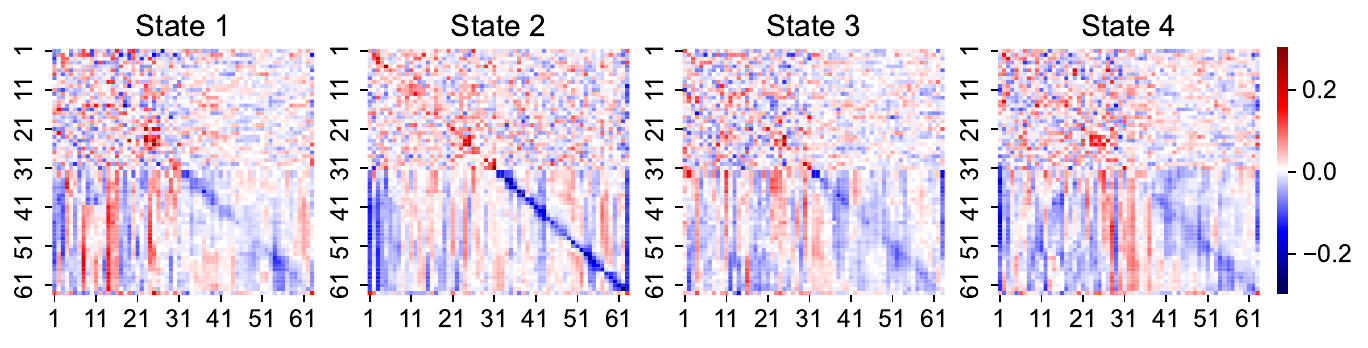}
\caption{Estimated coefficient matrices.}
\label{fig:coefficient}
\end{figure}

\begin{figure}[!ht]
\centering
\includegraphics[width = 0.75\textwidth]{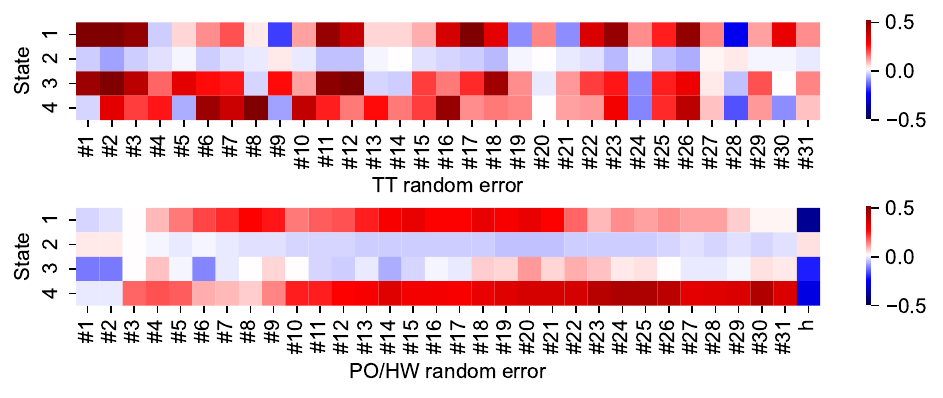}
\caption{Estimated mean vectors of the random error term.}
\label{fig:mean_vec}
\end{figure}

\begin{figure}[!ht]
\centering
\includegraphics[width = 0.8\textwidth]{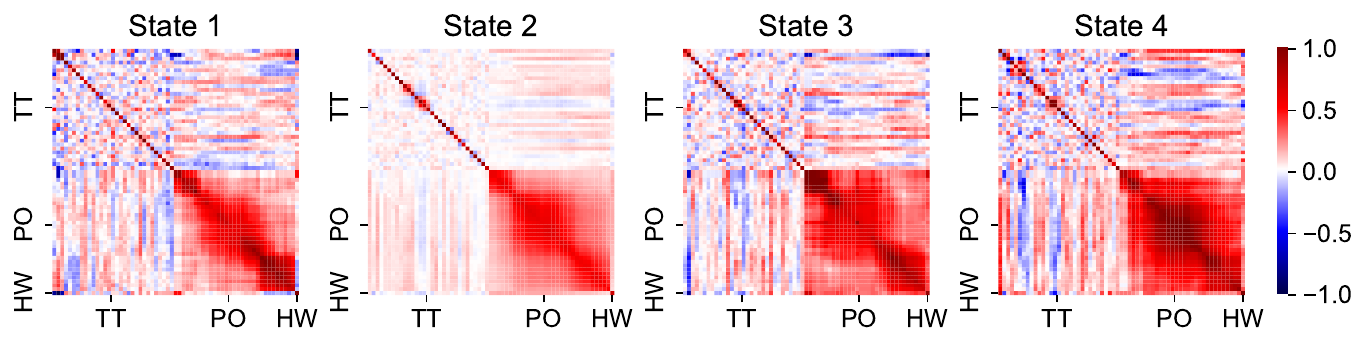}
\caption{Estimated covariance matrices of the random error term.}
\label{fig:cov_matrices}
\end{figure}

Bayesian models are powerful tools for interpreting parameters and uncovering patterns in probabilistic forecasting of bus travel time and passenger occupancy. In our study, as depicted in Fig.~\ref{fig:transition}, we showcase the estimated transition matrix, a cornerstone of the Markov regime-switching model. Each element of this matrix provides insight into the probability of transitioning from one state to another. The structure and values within this matrix are instrumental in understanding how frequently and likely certain state transitions occur, which in turn, can be linked to specific conditions or patterns in bus travel time and occupancy. We can see that most buses would like to keep or transit to state 2. Furthermore, Fig.~\ref{fig:coefficient} presents the estimated coefficient matrices of different states and we can find that they have different patterns, indicating that different states show different relationships between adjacent buses. Fig.~\ref{fig:mean_vec} and Fig.~\ref{fig:cov_matrices} illustrate the estimated mean vectors and covariance matrices of the random error term. The visual representation of mean vectors gives us an understanding of the central tendencies of errors across different states. The covariance matrices, on the other hand, unravel the relationships between errors of different variables. These visualizations reveal clear distinctions among different states in both mean values and covariance matrices.

\subsection{Forecast distribution}
\begin{figure}[!ht]
\centering
\includegraphics[width = 0.9\textwidth]{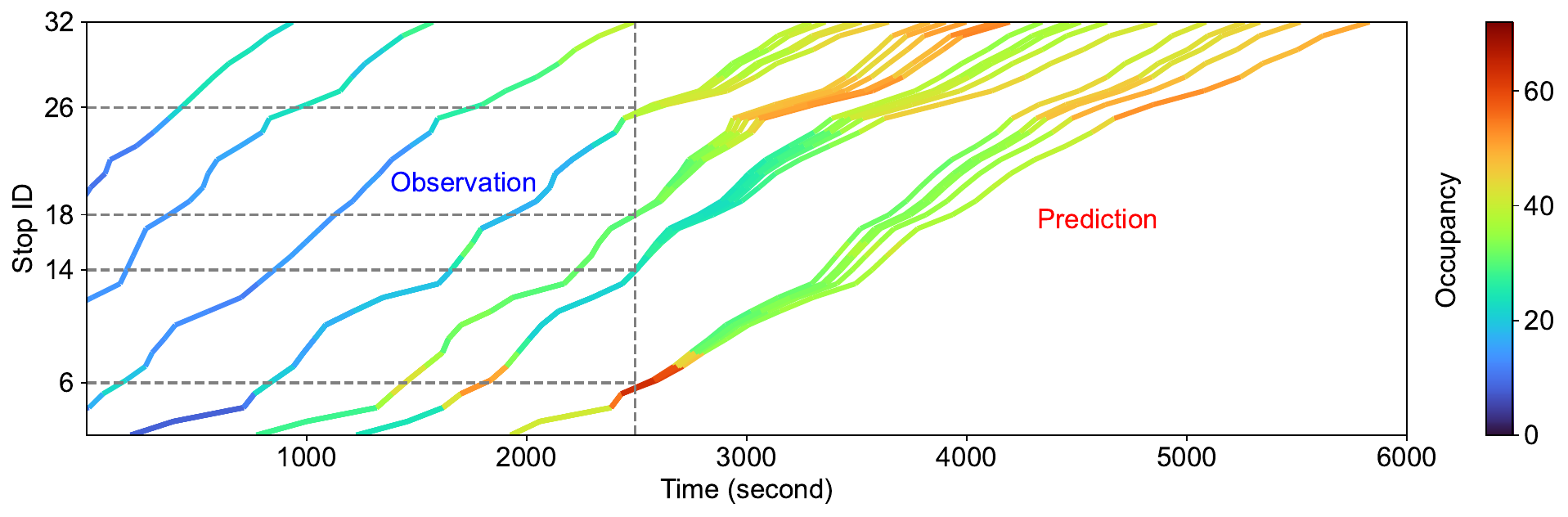}
\caption{Trajectory plot to show the bus travel time and occupancy forecasting. The vertical line represents the current time that separates past and future. Each colored curve shows the trajectory of one bus run, with color indicating the number of  passengers onboard (i.e., occupancy). For each bus, we plot five samples of the predicted downstream travel time and passenger occupancy based on the proposed model.}
\label{fig:intro_problem}
\end{figure}

Figure~\ref{fig:intro_problem} shows that the last bus has arrived at stop \#14 and the goal is to provide predicted distributions for travel time and passenger occupancy of downstream links (that is, from stop \#14 to \#32), and provide trip travel time/arrival time distributions. In this figure, we plot the predicted trajectories with passenger occupancy. In particular, the model generates multiple outcomes for each bus run, depicted by the spread of five sampled trajectories, which collectively offer a distribution that encapsulates the possible variance in travel times and occupancy. This spread of predictions signifies the model's robust approach to capturing the uncertainties inherent in the bus systems.

\begin{figure}[!ht]
\centering
\subfigure[Forecasting probability distribution of link travel time.]{
    \centering
    \includegraphics[width=0.65
    \textwidth]{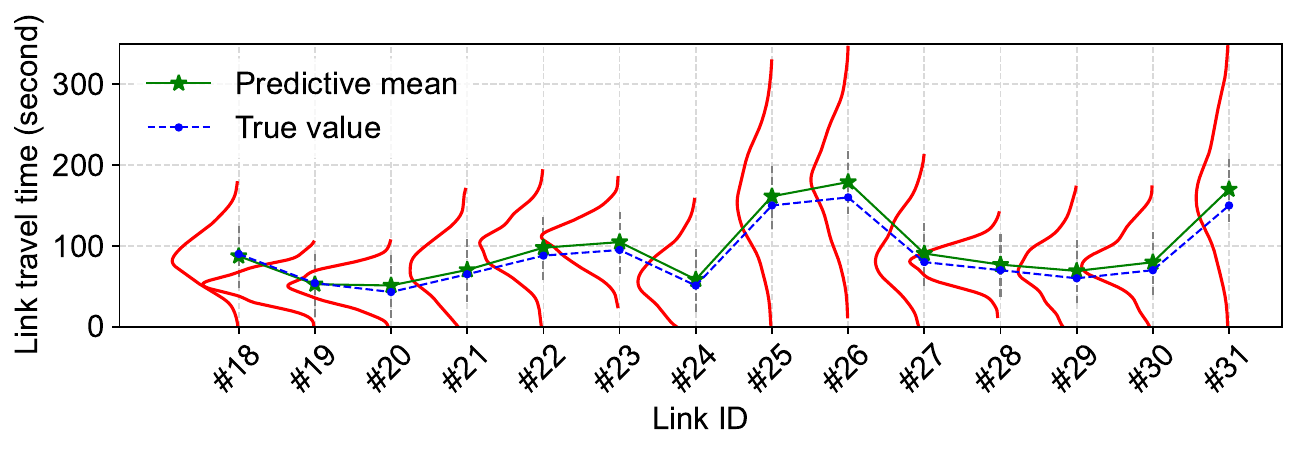}
}
\subfigure[Forecasting probability distribution of link passenger occupancy.]{
    \centering
    \includegraphics[width=0.65
    \textwidth]{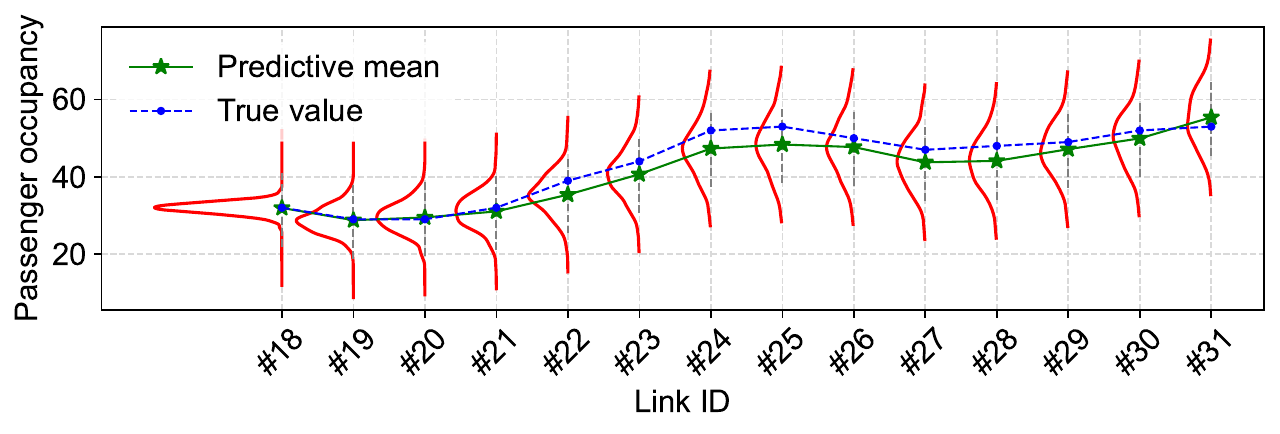}
}
\subfigure[Forecasting probability distribution of trip travel time.]{
    \centering
    \includegraphics[width=0.65
    \textwidth]{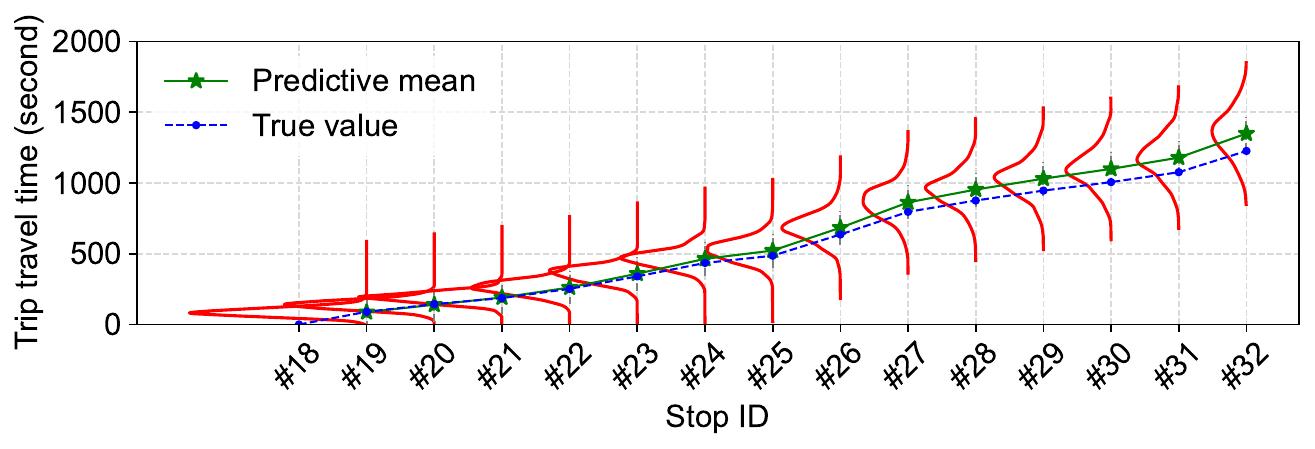}
}
\caption{Probability distributions of forecasting travel time and passenger occupancy.}
\label{fig:4cast_dist}
\end{figure}

We present visualizations of predicted probability distributions for a specific bus at a particular time point. Using a sampling method, obtaining the predicted trip travel time for the bus becomes straightforward. The visualization encompasses forecasting probability distributions for link travel time, passenger occupancy, and trip travel time, as demonstrated in Fig.~\ref{fig:4cast_dist}. In this figure, the bus has already traversed the first 17 links and our goal is to forecast the next 14 links (from \#18 to \#31). The blue points represent the true values, while the green points represent the predictive mean values. The first two panels show the predicted probability distributions for link travel times and passenger occupancy. Evidently, the predictive means closely align with the true values, which confirms the good accuracy of our forecasting for both bus link travel time and passenger occupancy. Furthermore, we observe that link travel times with larger values tend to exhibit larger variances, indicated by the larger variance in those red density functions. Additionally, upcoming links situated near the current links have smaller variances, suggesting that more precise predictions. The bottom panel shows the forecasting probability distributions for trip travel times. As the number of links in a trip increases, we notice that the red bell curves become more spread out, reflecting an increased variance in trip travel time. This is because longer trips inherently introduce more uncertainty in travel time predictions.

\section{Discussion}\label{sec:con}
In this paper, we propose a Bayesian Markov regime-switching vector autoregressive model for probabilistic forecasting of bus travel time and passenger occupancy. Our approach can effectively capture and address several critical factors: the correlations between travel time and passenger occupancy, the relationship between adjacent buses, and the multimodality/skewness of bus travel time and passenger occupancy distributions. To validate our proposed model, we conduct extensive numerical experiments on a real-world dataset. Our results demonstrate the superiority of the proposed approach compared to benchmark models and its effectiveness in providing accurate forecasts for bus travel time and passenger occupancy.

Our approach has implications for both practice and research. First, the proposed Bayesian Markov regime-switching vector autoregressive model could forecast trip travel time (i.e., estimated time of arrival) and passenger occupancy distributions, which could be incorporated into real-time bus information systems to help passengers and bus agencies make better decisions. Second, the proposed Bayesian model is also an interpretable tool for bus agencies to better understand bus operation patterns with uncertainty. For example, if one has access to enormous historical bus operation data (including some special events such as extreme weather, sporting events, large-scale concerts, etc.), we could also learn the pattern of special events, which would be helpful for bus agencies to provide better and robust management and operations. Third, our model can also be used to model other transport systems with interactions between adjacent vehicles. For instance, we can use the same model to model train/metro operation in a network and study how delay propagates. The proposed Bayesian Markov regime-switching vector autoregressive model can offer valuable insights into understanding and mitigating the delays that frequently affect train systems. Last, the proposed Bayesian Markov regime-switching vector autoregressive model could be utilized to perform imputation for series data with missing values. 

\section*{Acknowledgements}

This research is supported in part by the Canadian Statistical Sciences Institute (CANSSI) Collaborative Research Teams (CRT) grants and in part by the Natural Sciences and Engineering Research Council (NSERC) of Canada. X. Chen acknowledges the funding support from the China Scholarship Council (CSC).

\bibliographystyle{elsarticle-harv}
\bibliography{references}
\pagebreak
\appendix
\section{Gibbs sampling algorithm for parameter estimation.}\label{appA}
\begin{algorithm}[H]
\renewcommand{\algorithmicrequire}{\textbf{Input:}}
\renewcommand{\algorithmicensure}{\textbf{Output:}}
\caption{Gibbs sampling for parameter estimation.}\label{alg:gibbs}
\begin{algorithmic}[1]
{\fontsize{10pt}{18pt}\selectfont
\REQUIRE Sequential observations $\left\{\boldsymbol{y}_i^{\left(d\right)} \right\}_{i=1,d=1}^{I_d,D}$, hyperparameters $\Theta$ and $\boldsymbol{\alpha}$, random initialization of sequential states $\left\{z_i^{\left(d\right)} \right\}_{i=1,d=1}^{I_d,D}$, iterations $n_1$, $n_2$.
\ENSURE Samples of transition matrix $\left\{{\boldsymbol{\pi}_k}^{(\rho)}\right\}_{k=1,\rho=1}^{K,n_2}$, samples of mean vectors $\left\{\boldsymbol{\mu}_k^{(\rho)}\right\}_{k=1,\rho=1}^{K,n_2}$, and samples of covariance matrices $\left\{\boldsymbol{\Sigma}^{(\rho)}_k\right\}_{k=1,\rho=1}^{K,n_2}$.
\FOR{$\mathrm{iter}=1$ to $n_1+n_2$}
\FOR{$k=1$ to $K$}
\STATE Draw $\boldsymbol{\Sigma}_k$ and $\boldsymbol{\mu}_k$ according to Eq.~\eqref{eq:cov prior} and Eq.~\eqref{eq:mean prior}.
\STATE Draw $\boldsymbol{A}_k$ according to Eq.~\eqref{eq:A prior}.
\IF{$\mathrm{iter}>n_1$}
    \STATE Collect $\boldsymbol{\mu}_k$, $\boldsymbol{\Sigma}_k$, and $\boldsymbol{A}_k$ to the output sets.
\ENDIF
\ENDFOR
\FOR{$k=1$ to $K$}
\STATE Draw $\boldsymbol{\pi}_k$ according to Eq.~\eqref{eq:pi prior}.
\IF{$\mathrm{iter}>n_1$}
\STATE Collect $\boldsymbol{\pi}_k$ to the output set.
\ENDIF
\ENDFOR
\STATE Conduct the forward-backward algorithm to obtain $\alpha\left(\cdot\right)$ and $\beta\left(\cdot\right)$.
\FOR{$d=1$ to $T$}
\STATE Calculate $\boldsymbol{\pi}^*$ and draw $z_1^{\left(d\right)}$ according to Eq.~\eqref{sample_z1}.
\FOR{$i=2$ to $I_d$}
    \STATE Calculate $p\left(z_i^{\left(d\right)}\right)$ according to Eq.~\eqref{eq:fb}.
    \STATE Draw $z_i^{\left(d\right)}$ according to Eq.~\eqref{eq:zpdf}.
\ENDFOR
\ENDFOR
\FOR{$k=1$ to $K$}
\STATE Update the parameters $\Theta=\{\boldsymbol{\mu}_0,\lambda_0, \boldsymbol{\Psi}_0, \nu_0\}$ by Eq.~\eqref{l123}.
\ENDFOR
\STATE Update the parameters $\boldsymbol{\alpha}$ by Eq.~\eqref{sample_pi}.
\ENDFOR
\RETURN{$\left\{{\boldsymbol{\pi}_k}^{(\rho)}\right\}_{k=1,\rho=1}^{K,n_2}$,$\left\{\boldsymbol{\mu}^{(\rho)}_k\right\}_{k=1,\rho=1}^{K,n_2}$, $\left\{\boldsymbol{\Sigma}^{(\rho)}_k\right\}_{k=1,\rho=1}^{K,n_2}$, $\left\{\boldsymbol{A}^{(\rho)}_k\right\}_{k=1,\rho=1}^{K,n_2}$.}
}
\end{algorithmic}
\end{algorithm}
\pagebreak
\section{Gibbs sampling algorithm for probabilistic forecasting.}\label{appB}

\begin{algorithm}[H]
\renewcommand{\algorithmicrequire}{\textbf{Input:}}
\renewcommand{\algorithmicensure}{\textbf{Output:}}
\caption{Gibbs sampling for probabilistic forecasting.}\label{alg:4cast}
\begin{algorithmic}[1]
{\fontsize{10pt}{20pt}\selectfont
\REQUIRE Sequential observations $\left\{\boldsymbol{y}_1,\dots,\boldsymbol{y}_{j-1},\boldsymbol{y}_j^o,\boldsymbol{y}_{j+1}^o,\dots,\boldsymbol{y}_J^o\right\}$, samples of transition matrix $\left\{{\boldsymbol{\pi}_k}^{(\rho)}\right\}_{k=1,\rho=1}^{K,n_2}$, samples of mean vectors $\left\{\boldsymbol{\mu}_k^{(\rho)}\right\}_{k=1,\rho=1}^{K,n_2}$, samples of covariance matrices $\left\{\boldsymbol{\Sigma}^{(\rho)}_k\right\}_{k=1,\rho=1}^{K,n_2}$, and samples of coefficient matrices $\left\{\boldsymbol{A}^{(\rho)}_k\right\}_{k=1,\rho=1}^{K,n_2}$.
\ENSURE Forecasting sample set $\left\{{\boldsymbol{y}_j^f}^{(\rho)},\dots,{\boldsymbol{y}_J^f}^{(\rho)}\right\}_{\rho=1}^{n_2}$.
\FOR{$\rho=1$ to $n_2$}
\FOR{$j^\prime=j$ to $J$}
\STATE Draw $\boldsymbol{y}_{j^\prime}^f$ as the forecasting sample.
    \STATE Collect $\boldsymbol{y}_{j^\prime}^f$ to the forecasting sample set.
\ENDFOR
\STATE Calculate $\boldsymbol{\pi}^*$ and draw $z_1$ according to Eq.~\eqref{sample_z1}.
\STATE Conduct the forward-backward algorithm to obtain $\alpha\left(\cdot\right)$ and $\beta\left(\cdot\right)$.
\FOR{$j=2$ to $J$}
    \STATE Calculate $p\left(z_j\right)$ according to Eq.~\eqref{eq:fb}.
    \STATE Draw $z_j$ according to Eq.~\eqref{eq:zpdf}.
\ENDFOR
\ENDFOR
\RETURN{$\left\{{\boldsymbol{y}_j^f}^{(\rho)},\dots,{\boldsymbol{y}_J^f}^{(\rho)}\right\}_{\rho=1}^{n_2}$.}
}
\end{algorithmic}
\end{algorithm}

\end{document}